\documentclass[aps,pra,twocolumn,superscriptaddress,amsmath,amssymb,showpacs]{revtex4-1}
\pdfoutput=1
\usepackage[caption=false]{subfig}
\usepackage{algorithm}
\usepackage{color}
\usepackage{algorithmic}
\usepackage{amsmath}
\usepackage{amssymb}
\usepackage{amsthm}
\usepackage{bm}
\usepackage{thmtools}

\usepackage{array}
\usepackage{url}
\usepackage{hyperref}
\usepackage{graphicx}
\usepackage{bibentry}
\usepackage{microtype}

\providecommand{\U}[1]{\protect\rule{.1in}{.1in}}
\newtheorem*{theorem*}{Theorem}

\newtheorem{result}{Result}

\begin{document}
\title{Covert sensing using floodlight illumination}
\author{Christos N. Gagatsos}
\affiliation{College of Optical Sciences, University of Arizona, 1630 E. University Blvd., Tucson, Arizona 85719, USA}
\author{Boulat A. Bash}
\affiliation{Department of Electrical and Computer Engineering, University of Arizona, 1230 E Speedway Blvd., Tucson, Arizona 85719, USA}
\author{Animesh Datta}
\affiliation{Department of Physics, University of Warwick, Coventry, CV4 7AL, UK}
\author{Zheshen Zhang}
\affiliation{Department of Materials Science and Engineering, University of Arizona, 1235 E. James E. Rogers Way, Tucson, Arizona 85721, USA}
\affiliation{College of Optical Sciences, University of Arizona, 1630 E. University Blvd., Tucson, Arizona 85719, USA}
\author{Saikat Guha}
\affiliation{College of Optical Sciences, University of Arizona, 1630 E. University Blvd., Tucson, Arizona 85719, USA}
\affiliation{Department of Electrical and Computer Engineering, University of Arizona, 1230 E Speedway Blvd., Tucson, Arizona 85719, USA}

\begin{abstract}
We propose a scheme for covert active sensing using floodlight illumination from a THz-bandwidth amplified spontaneous emission (ASE) source and heterodyne detection. We evaluate the quantum-estimation-theoretic performance limit of covert sensing, wherein a transmitter's attempt to sense a target phase is kept undetectable to a quantum-equipped passive adversary, by hiding the signal photons under the thermal noise floor. Despite the quantum state of each mode of the ASE source being mixed (thermal), and hence inferior compared to the pure coherent state of a laser mode, the thousand-times higher optical bandwidth of the ASE source results in achieving a substantially superior performance compared to a narrowband laser source by allowing the probe light to be spread over many more orthogonal temporal modes within a given integration time. Even though our analysis is restricted to single-mode phase sensing, this system could be applicable extendible for various practical optical sensing applications.
\end{abstract}
\maketitle

\indent
\emph{Introduction.}~Quantum technologies and in particular quantum and quantum-enhanced sensing \cite{Giov2006} have high-impact near-term applications that include microscopy \cite{micro1,micro2}, vibrometry \cite{Rembe2017}, ranging \cite{Rarity1990}, astronomy \cite{Mankei2017}, and medical imaging \cite{Gavin2017}. Moreover, they have contributed to fundamental discoveries such as gravitational waves detection \cite{GW1}. In particular, the problem of high-precision estimation of a small unknown phase modulation has attracted a lot of attention in the quantum sensing literature, because of its applicability to many of the aforesaid real-life optical imaging and sensing problems. The quantum-physical modeling of a typical sensor results in the quantum state $\rho(\theta)$ of a single mode of the target-modulated light encoding a parameter of interest $\theta$. We will be interested in scenarios where $n \approx WT$ independent temporal modes are modulated identically by a static target resulting in a target-return state of the form, $\rho(\theta)^{\otimes n}$, where $W$ (Hz) is the optical bandwidth of the optical source, and $T$ (s) is the interrogation time. One desires to find the optimal measurement on this target-return light that generates an estimate $\hat{\theta}_n$ with the minimum possible deviation from the true value, often quantified as the variance, or the mean squared error (MSE), $\langle (\theta-\hat{\theta}_n)^2$. The quantum Cram\'er-Rao bound (QCRB) \cite{demkowicz15quantmetrologysurvey} gives a lower bound on the MSE attainable over all measurements generating an unbiased estimate of $\theta$, which is given by: $\langle (\theta-\hat{\theta}_n)^2\rangle\geq1/(nF)$, where $F$ is the quantum Fisher information (QFI) of $\theta$, calculated on $\rho(\theta)$. The bound gets asymptotically tight as $n \to \infty$. In this Letter, we discuss the notion of quantum-secure covert sensing, inspired by recent work on covert communications \cite{Bash2012,che13sqrtlawbsc,wang15covert,bash15covertbosoniccomm}. The task is to minimize the MSE of sensing an unknown carrier-phase modulation $\theta$ by using active illumination, while ensuring that an all-powerful non-line-of-sight passive adversary Willie who is able to observe the entire electromagnetic background around the sensor's field of operation equipped with all measurements allowed by quantum physics, and who has knowledge of the exact probing interval and the true value of $\theta$, cannot reliably distinguish between the two equally-likely hypotheses: $H_0$ (the target is not being probed), and $H_1$ (the target is being probed). A sensing scheme is $\epsilon$-covert if we can ensure Willie's detection error probability to satisfy, $1/2 - \epsilon \le {\mathbb P}_e^{(\text{det})} \le 1/2,\ \epsilon>0$, i.e., his ability to distinguish the two hypotheses is $\epsilon$-close to that of flipping a coin. Our goal is to find---with the optimal choice of an optical transmitter and receiver---the minimum MSE that can be attained as a function of $n$ (which is proportional to the integration time $T$ for a given optical bandwidth $W$), while ensuring $\epsilon$-covert operation. We note that passive sensing---i.e., sensing a self-luminous or naturally-illuminated scene---is by definition the most covert form of sensing. The quintessential example of passive sensing is human vision. Passive sensing can however be impractical when the illumination levels are so weak that the signal-to-noise ratio at the receiver is insufficient to obtain the desired accuracy within the integration time, or if the target is hidden from direct sight. In this paper, we focus on covert sensing using active illumination \cite{PatentBash2018} and consider the problem of phase estimation in the presence of loss and noise.
\begin{figure}
	\centering
\includegraphics[width=1\linewidth]{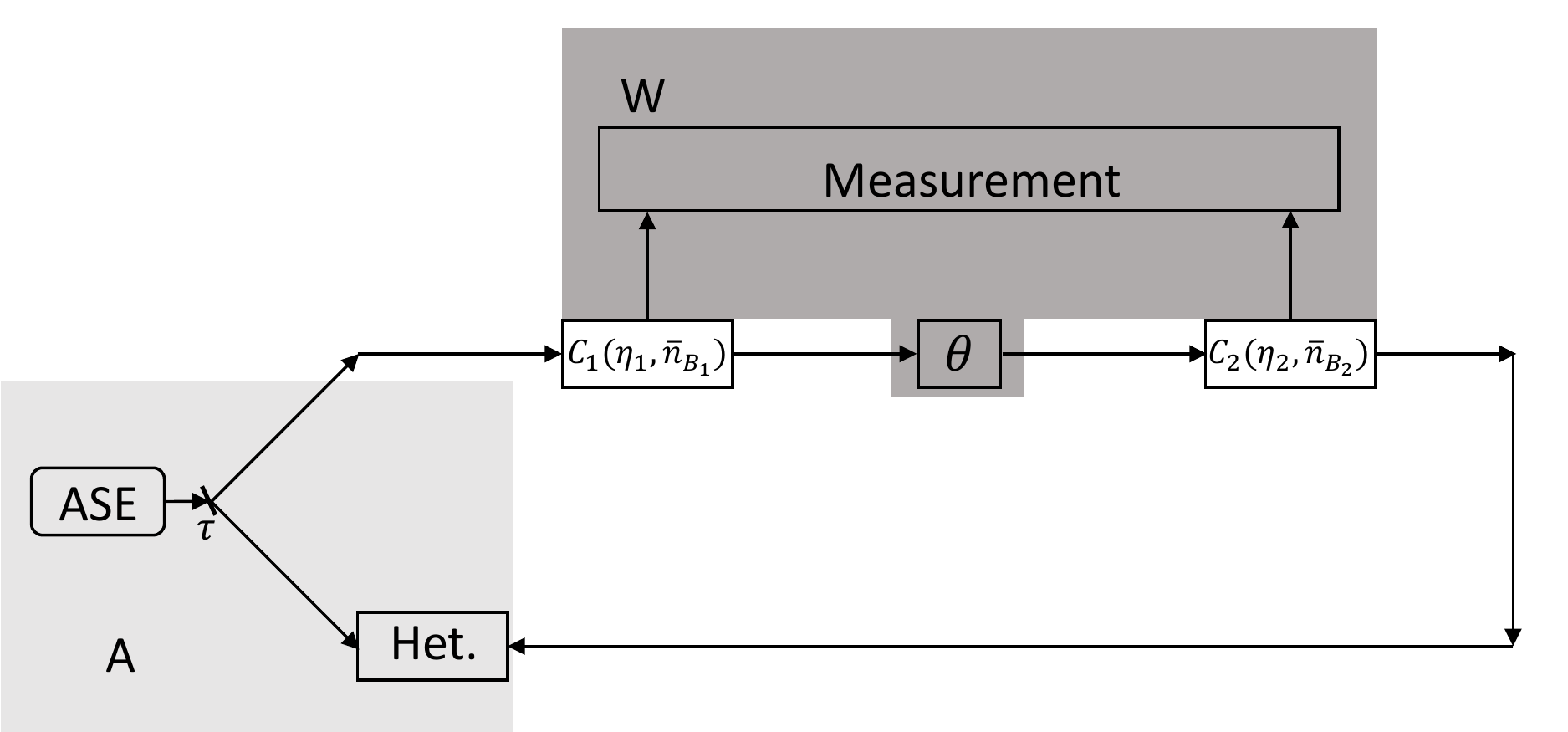}
	\caption{Alice splits the output of the source on a highly-unbalanced beam splitter with controllable transimissivity $\tau$. She sends the weaker of the two beams along a forward propagation path, a single-mode bosonic channel $C_1(\eta_1, {\bar n}_{B_1})$ of transmissivity $\eta_1$ and additive thermal noise of mean photon number ${\bar n}_{B_1}$ per mode. The propagated light, after accruing an unknown target phase $\theta$ propagates back to Alice on a single-mode bosonic channel $C_1(\eta_2, {\bar n}_{B_2})$. The brighter of the two beams at the transmitter is held locally for use as local oscillator to heterodyne detect the target-return light. All the lost photons on both the forward and return path channels are given to Willie, who is allowed to make an arbitrary quantum measurement.}
	\label{fig:thermal2}
\end{figure}\\
\indent \emph{Problem setup.}~Figure \ref{fig:thermal2} is a schematic of the problem setup. Alice (the sensor) generates an $n$-mode optical probe, each mode of which traverses the forward propagation path to the target, modeled by a single-mode bosonic channel $C_1(\eta_1, {\bar n}_{B_1})$ of transmissivity $\eta_1$ and additive thermal noise of mean photon number ${\bar n}_{B_1}$ per mode. Each mode of the output light illuminates the same (unknown) target phase $\theta$. The target-modulated light propagates through the return channel $C_2(\eta_2, {\bar n}_{B_2})$ back to Alice. All the lost photons on both the forward and return path channels are given to Willie, who is allowed to make an arbitrary measurement on this $2n$-mode optical state in order to do his hypothesis test to determine whether Alice is sensing or not. The problem is to quantify the minimum MSE attainable as a function of $n$, with optimal choices of Alice's probe state, receiver, and the transmitted mean photon number per mode ${\bar n}_S$ so as to ensure $\epsilon$-covertness no matter what measurement Willie uses. We assume the forward and return path transmissivities $\eta_1$ and $\eta_2$ to be known to Alice, and different in general, to account for bi-static sensor configurations with unequal path lengths. We also allow for different thermal noise temperatures for the forward and return path channels, to account for scenarios such as one of the two paths being actively jammed. 

\begin{figure}
	\centering
	\includegraphics[width=1\linewidth]{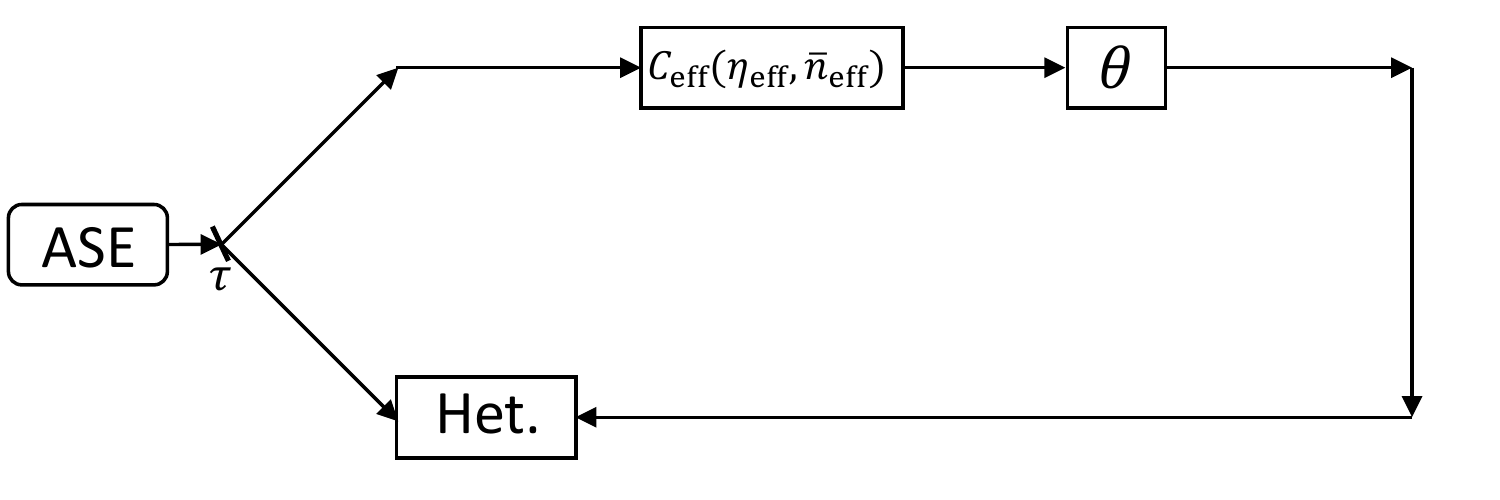}
	\caption{For the setup of Fig.~\ref{fig:thermal2}, Alice sees one effective thermal channel $C_{\textrm{eff}}$. This setup was used in \cite{bash2017isit} for covert sensing with a coherent state probe.}
	\label{fig:thermal1}
\end{figure}

\emph{Main results.}~For our \emph{achievability} results, we will consider two sensor probes: (1) a laser-light (coherent-state) transmitter and heterodyne detection, and (2) a broadband amplified spontaneous emission (ASE) noise source (see Fig.~\ref{fig:thermal2}), whose output is split on a highly unbalanced beam splitter which produces a phase-insensitive classically-correlated two-mode output. The weaker of the two outputs is used to probe the target, while the brighter beam is kept locally at the transmitter to be used as local oscillator to heterodyne detect the target-return light. Our two main results are:
\begin{result}
Alice can achieve an MSE, $\langle (\theta - \hat{\theta}_n)^2 \rangle = \mathcal{O}(1/\sqrt{n})$. Attempting to have the MSE diminish any faster as a function of $n$ must result in detection by Willie with high probability. 
\end{result}
\begin{result}
The MSE when Alice uses a coherent state probe state scales as $\langle (\theta - \hat{\theta}_n)^2 \rangle = c_{\textrm{coh}}/(\epsilon\sqrt{n}),\ \epsilon>0$, while when she uses the ASE source $\langle (\theta - \hat{\theta}_n)^2 \rangle = c_{\textrm{ASE}}/(\epsilon\sqrt{n}),\ \epsilon>0$. The per-mode performance of the coherent state probe is better than the ASE source probe, i.e., $c_{\textrm{coh}}\leq c_{\textrm{ASE}}$. However, the ASE source can achieve a much lower MSE for a given integration time $T$, since $n \approx WT$ and optical bandwidth $W$ of the ASE source is larger by $2$-$3$ orders of magnitude.
\end{result}
\indent \emph{Covert sensing with a coherent state probe.}~Let us summarize the main results from \cite{bash2017isit}. The model is depicted in Fig.~\ref{fig:thermal1} with $\eta_{\textrm{eff}}=\eta$ and $\bar{n}_{\textrm{B}_{\textrm{eff}}} = \bar{n}_{\textrm{B}}$. Over $n$ modes, Alice transmits a mean photon number of $\langle N_{\textrm{S}} \rangle=n \bar{n}_{\textrm{S}}$, where $\bar{n}_{\textrm{S}}$ is the mean transmitted photon number per mode. We make the assumption that the total variance, $\langle \Delta N_{\textrm{S}}^2 \rangle=\mathcal{O}(n)$. Let us begin by stating the \emph{converse}: Alice's sensing attempt is either detected by Willie with arbitrarily low detection probability, or the MSE of Alice's estimation is tightly lower bounded by $1/\sqrt{n}$, i.e.,  $\langle (\theta - \hat{\theta}_n)^2 \rangle = \Omega(1/\sqrt{n})$ \cite{bash2017isit}. This notation implies that if Alice tries to get her MSE to surpass the $1/\sqrt{n}$ scaling (e.g., by increasing her transmit power), Willie will detect her sensing attempt with high probability. Now, we state the \emph{achievability}: Suppose that Willie can perform any measurement permitted by physics on the $2n$ optical modes he collects. Alice can lower bound Willie's detection error probability $\mathbb{P}_e^{(\textrm{det})}\geq 1/2-\epsilon$, for any $\epsilon>0$ while achieving the MSE $\langle (\theta - \hat{\theta}_n)^2 \rangle = \mathcal{O}(1/\sqrt{n})$ using an $n$-mode coherent-state probe. For a coherent probe state $\otimes_{i=1}^n|\alpha_i\rangle$, where each complex amplitude $\alpha_i$ is drawn from the distribution $p(\alpha)=e^{-|\alpha|^2/\bar{n}_{\textrm{S}}}/\pi \bar{n}_{\textrm{S}}$, Willie's detection error probability is shown to satisfy \cite{bash2017isit}:
\begin{eqnarray}
\nonumber \mathbb{P}_e^{(\textrm{det})}\geq \frac{1}{2}-\frac{(1-\eta)\bar{n}_{\textrm{S}}\sqrt{n}}{4\sqrt{\eta \bar{n}_{\textrm{B}}(1+\eta)}}.
\end{eqnarray}
Alice can choose
\begin{eqnarray}
\nonumber \bar{n}_{\textrm{S}} = \frac{4\epsilon \sqrt{\eta \bar{n}_{\textrm{B}}(1+\eta \bar{n}_{\textrm{B}})}}{\sqrt{n}(1-\eta)}
\end{eqnarray}
to ensure $\epsilon$-covertness \cite{bash2017isit}. Using the aforesaid coherent state probe and ideal heterodyne detection measurement, Alice can achieve the MSE \cite{bash2017isit}:
\begin{eqnarray}
\nonumber \langle (\theta - \hat{\theta}_n)^2 \rangle \approx \frac{c_{\textrm{het}}}{\epsilon\sqrt{n}},
\end{eqnarray}
where
\begin{eqnarray}
\nonumber c_{\textrm{het}} = \frac{(1-\eta)(1+\bar{n}_{\textrm{B}}(1-\eta))}{8 \eta \sqrt{\eta \bar{n}_{\textrm{B}}(1+\eta \bar{n}_{\textrm{B}})}}.
\end{eqnarray}
Using a coherent state probe, but no restriction on the receiver, the MSE achieved is at best two times lower than that is achieved with a heterodyne receiver \cite{bash2017isit}. In other words, with Alice's optimal receiver,
\begin{eqnarray}
\nonumber \langle (\theta - \hat{\theta}_n)^2 \rangle \geq \frac{c_{\textrm{coh}}}{\epsilon\sqrt{n}},
\end{eqnarray}
where
\begin{eqnarray}
\nonumber c_{\textrm{coh}} = \frac{(1-\eta)(1+2\bar{n}_\textrm{B}(1-\eta))}{16 \eta \sqrt{\eta \bar{n}_{\textrm{B}}(1+\eta \bar{n}_B)}}.
\end{eqnarray}
\indent\emph{Covert sensing with ASE floodlight illumination.}~Let us first consider the \emph{achievability}. Since all states and channels involved are Gaussian, we will work with the symplectic formalism. Further, all the first moments of the states involved are zero. The ASE source, after being split on an un-balanced beam-splitter, emits a two-mode zero-mean Gaussian state with covariance matrix (CM) given by \cite{zhuang15floodlightQKD}:
\begin{eqnarray}
V_{\textrm{ASE}} = 
\begin{pmatrix}
A_{\textrm{ASE}} & 0_{2\times 2} \\
0_{2\times 2} & A_{\textrm{ASE}}
\end{pmatrix},
\end{eqnarray}
where
\begin{eqnarray}
A_{\textrm{ASE}} = 
\begin{pmatrix}
\bar{n}_{\textrm{S}}+\frac{1}{2} & \sqrt{\bar{n}_{\textrm{S}} \bar{n}_{\textrm{LO}}}\\
\sqrt{\bar{n}_{\textrm{S}} \bar{n}_{\textrm{LO}}} & \bar{n}_{\textrm{LO}}+\frac{1}{2}
\end{pmatrix}.
\end{eqnarray}
Here, $\bar{n}_{\textrm{S}}$ is the mean transmit photon number per mode, and $\bar{n}_{\textrm{LO}} \gg \bar{n}_{\textrm{S}}$ is the mean photon number per mode of Alice's local oscillator (LO) for her heterodyne detection receiver. If Alice is probing the target, Willie gets a state with the following CM (see App. Sec.\ref{SM:CM}),
\begin{eqnarray}
V_W =
\begin{pmatrix}
B_W & C_W \\
C_W^T & B_W
\end{pmatrix},
\label{VW}
\end{eqnarray}
where
\begin{eqnarray}
B_W =
\begin{pmatrix}
	w_{11} & -w_{12}\cos\theta\\
   -w_{12}\cos\theta & w_{22}
\end{pmatrix},\,{\text{and}}
\end{eqnarray}
\begin{eqnarray}
C_W =
\begin{pmatrix}
0 & w_{12} \sin\theta \\
-w_{12}\sin\theta & 0
\end{pmatrix},
\end{eqnarray}
where $w_{11} = (1-\eta_2)\eta_1 \bar{n}_{\textrm{S}} + (1-\eta_1)(1-\eta_2)\bar{n}_{\textrm{B}_1}+\eta_2 \bar{n}_{\textrm{B}_2} + 1/2,\ 
w_{12} = \sqrt{(1-\eta_2)\eta_1(1-\eta_1)} (\bar{n}_{B_1}-\bar{n}_S),\ w_{22} =  \eta_1 \bar{n}_{B_1} + (1-\eta_1) \bar{n}_\textrm{S} + 1/2.$
When Alice is not sensing, Willie's CM is given by (see App. Sec. \ref{SM:CM}) $V_W^{(0)}=V_W(\bar{n}_{\textrm{S}}=0)$, where $w_{ij}^{(0)}=w_{ij}(\bar{n}_S=0)$. 
We assume $\bar{n}_S \ll \bar{n}_{\textrm{B}_1}$ and $\bar{n}_\textrm{S}\ll\bar{n}_{\textrm{B}_2}$, as this is intuitively needed for remaining covert. Willie must discriminate between the state $\hat{\rho}_1$ with CM $V_W$ (Alice is sensing) and the state $\hat{\rho}_0$ with CM $V_W^{(0)}$ (Alice is not sensing). After transmission of $n$ modes over the probing interval, Willie's average probability of discrimination error is
\begin{align*}
\mathbb{P}_{e}^{(\mathrm{w})}&\geq\frac12\left[1 - \frac12\| \hat{\rho}_0^{\otimes n} - \hat{\rho}_1^{\otimes n} \|_1\right].
\end{align*}
See also App. Sec. \ref{SM:Detect} for a brief discussion on detectability.
The trace distance $\|\hat{\rho}_0-\hat{\rho}_1\|_1$ between states 
$\hat{\rho}_0$ and $\hat{\rho}_1$ is upper-bounded by the quantum relative entropy (QRE) $D(\hat{\rho}_0\|\hat{\rho}_1)=\textrm{tr}\left(\hat{\rho}_0\ln \hat{\rho}_0 \right)-\textrm{tr}\left(\hat{\rho}_0\ln \hat{\rho}_1 \right)$ using quantum Pinsker's inequality, i.e.,
$\|\hat{\rho}_0-\hat{\rho}_1\|_1\leq\sqrt{2D(\hat{\rho}_0\|\hat{\rho}_1)}$
which implies that
\begin{align}
\label{eq:qre_pe_lb}\mathbb{P}_{e}^{(\mathrm{w})}&\geq\frac{1}{2}-\sqrt{\frac{1}{8}D(\hat{\rho}_0^{\otimes n}\|\hat{\rho}_1^{\otimes n})}.
\end{align}
Thus, ensuring that $D(\hat{\rho}_0^{\otimes n}\|\hat{\rho}_1^{\otimes n})\leq 8\epsilon^2$ ensures that $\mathbb{P}_{e}^{(\mathrm{w})}\geq\frac{1}{2}-\epsilon$ over $n$ channel uses. QRE is additive for tensor product states, therefore
$D(\hat{\rho}_0^{\otimes n}\|\hat{\rho}_1^{\otimes n})=nD(\hat{\rho}_0\|\hat{\rho}_1).$
By expanding the QRE in Taylor series with respect to $\bar{n}_{\textrm{S}}$ at $\bar{n}_{\textrm{S}}=0$ and by using Taylor's theorem with the remainder we can upper bound $D(\hat{\rho}_0\|\hat{\rho}_1)$ by its second order term $c_2=f(\eta_1,\eta_2,\bar{n}_{\textrm{B}_1},\bar{n}_{\textrm{B}_2})>0$. For the aforesaid analysis on the QRE, see App. Sec. \ref{SM:Taylor}. Consequently Willie's error probability will be lower bounded as,
\begin{eqnarray}
\mathbb{P}_{e}^{(\mathrm{w})}\geq\frac{1}{2}-\frac{\sqrt{c_2}}{4}\sqrt{n}\,\bar{n}_S.
\label{eq:ProbErrorLB}
\end{eqnarray}
Thus, if Alice sets
\begin{eqnarray}
\label{covertns}\bar{n}_\textrm{S} =\frac{4 \epsilon}{\sqrt{c_2}}\frac{1}{\sqrt{n}},
\end{eqnarray}
she can ensure $\epsilon$-covertness, i.e., $\mathbb{P}_{e}^{(\mathrm{w})}\geq\frac{1}{2}-\epsilon$. As shown in Fig.~\ref{fig:thermal1}, Alice sees one effective thermal channel (see App. Sec. \ref{SM:Effective}), with $\eta_{\textrm{eff}}=\eta_1 \eta_2$ and $\bar{n}_{\textrm{B}_{\textrm{eff}}} = [\bar{n}_{\textrm{B}_1}(1-\eta_1)\eta_2+\bar{n}_{\textrm{B}_2}(1-\eta_2)]/(1-\eta_1 \eta_2)$, which simplifies the analysis on Alice's measurement. We consider heterodyne detection and in App. Sec. \ref{SM:Het} we prove that the heterodyne output comprises of a pair of quadrature measurements with Gaussian distributions $\mathcal{N}\left(\mu_i,\sigma_i^2\right),\ i=1,2$, of means $\mu_i$ and variances $\sigma_i^2$. $\mathcal{N}\left(\cos\theta,[1+\bar{n}_{\textrm{B}_\textrm{eff}}(1-\eta_\textrm{eff})]/(2\eta_{\textrm{eff}}\bar{n}_\textrm{S})+\cos^2\theta\right)$ and $\mathcal{N}\left(\sin\theta,[1+\bar{n}_{\textrm{B}_\textrm{eff}}(1-\eta_\textrm{eff})]/(2\eta_{\textrm{eff}}\bar{n}_\textrm{S})+\sin^2\theta\right)$.
The variances $\sigma_{1,2}^2$ are taken in the limit $\bar{n}_\textrm{LO}\rightarrow \infty$, i.e., a very bright local oscillator. By considering $n$ modes, taking into account that independent Gaussian variables are additive, substituting ${\bar n}_S$ from Eq. (\ref{covertns}) and assuming that $n \gg 1$, we get $\sigma_1^2 \approx \sigma_2^2=\sigma^2_{\textrm{het}}$ (App. Sec. \ref{SM:Het}) where
\begin{eqnarray}
\sigma^2_{\textrm{het}} = \frac{\tilde{c}_{\textrm{het}}}{\epsilon\sqrt{n}},
\end{eqnarray}
and $\tilde{c}_{\textrm{het}}=[1+\bar{n}_{\textrm{B}_\textrm{eff}}(1-\eta_\textrm{eff})]\sqrt{c_2}/(8\eta_{\textrm{eff}})$. Therefore,
\begin{eqnarray}
\label{MSEhet}\langle (\theta-\hat{\theta}_{\textrm{het},n})^2\rangle \approx \frac{\tilde{c}_{\textrm{het}}}{\epsilon\sqrt{n}}=\mathcal{O}\left(\frac{1}{\sqrt{n}}\right).
\end{eqnarray}
For a justification of the approximation in Eq. (\ref{MSEhet}), refer to \cite{bash2017isit} (Appendix C therein) and App. Sec. \ref{Sec:Estimator}. In the same limit $\bar{n}_\textrm{LO}\rightarrow \infty$, the QFI for the parameter $\theta$ for Alice is given by (see App. Sec. \ref{SM:QFI}):
\begin{eqnarray}
\label{qfiA} F_A = \frac{4 \bar{n}_{\textrm{S}} \eta_{\textrm{eff}} }{1+2\bar{n}_{\textrm{B}_\textrm{eff}}(1-\eta_\textrm{eff})}.
\end{eqnarray}
Therefore, by imposing the covertness condition Eq. \eqref{covertns} we get the QCRB for the ASE source,
\begin{eqnarray}
\label{MSEASE} \langle (\theta-\hat{\theta}_{n})^2\rangle \geq \frac{c_{\textrm{ASE}}}{\epsilon\sqrt{n}}
\end{eqnarray}
where $c_{\textrm{ASE}} = [1+2\bar{n}_{\textrm{B}_\textrm{eff}}(1-\eta_\textrm{eff})]\sqrt{c_2}/(16\eta_{\textrm{eff}}) $.
Therefore, with the ASE source, from Eqs. (\ref{MSEhet}) and (\ref{MSEASE}), heterodyne detection yields an MSE that is at most twice compared to what is attainable by the optimal quantum receiver.

Let us discuss briefly the \emph{converse}, i.e, the statement that regardless of the $n$-mode probe state employed by Alice, the sensing attempt is either detected with arbitrarily low detection probability, or the MSE of Alice's estimator is tightly lower bounded as  $\langle (\theta - \hat{\theta}_n)^2 \rangle = \Omega(1/\sqrt{n})$. Since for the converse a specific probe state is not fixed and having proved that Alice sees only one effective thermal channel with $(\eta_{\textrm{eff}},\ \bar{n}_{\textrm{B}_\textrm{eff}})$, the converse proved in \cite{bash2017isit} using the methods and results from \cite{Escher2011,Gagatsos2017bound} is still valid.\\
\indent \emph{Laser light vs. ASE source probe.}~Let us compare the quantum limits corresponding to the ASE and the coherent state sources. In \cite{bash2017isit}, it was found that for a coherent state probe, $\langle (\theta-\hat{\theta}_{n})^2\rangle \geq F_{\textrm{coh}}^{-1} =  {c_{\textrm{coh}}}/({\epsilon\sqrt{n}})$, where $c_{\textrm{coh}}=(1-\eta_\textrm{eff})[1+2\bar{n}_{\textrm{B}_\textrm{eff}}(1-\eta_\textrm{eff})]/[16\eta_{\textrm{eff}}\sqrt{\eta_{\textrm{eff}}\bar{n}_{\textrm{B}_\textrm{eff}}(1+\eta_{\textrm{eff}}\bar{n}_{\textrm{B}_\textrm{eff}})}]$ and $F_{\textrm{coh}}$ is the QFI for coherent state probe and for the effective single thermal channel Alice sees. When the ratio $\mu=F_{\textrm{A}}^{-1}/F_{\textrm{coh}}^{-1}$ is $\mu<1$, the ASE source can outperform the coherent state probe. For $n=WT$ channel uses, where the optical bandwidth $W$ depends on the source and probing time $T$ is fixed, the ratio is $\mu=\mu_c/\sqrt{\mu_w}$, where $\mu_c=c_{\textrm{ASE}}/c_{\textrm{coh}}$ and $\mu_w=W_{\textrm{ASE}}/W_{\textrm{coh}}$. 
If $\bar{n}_{\textrm{B}_1}=\bar{n}_{\textrm{B}_2}$, as expected in most situations since thermal equilibrium across the forward and return paths is not an onerous assumption, we get $\mu_c=1$ which is its minimum value (App. Sec. \ref{SM:Taylor}), i.e., the per-mode performance for the two sources is identical. Therefore, we have $\mu=\sqrt{W_{\textrm{coh}}/W_{\textrm{ASE}}}<1$ since the bandwidth of a typical laser source is in the GHz regime while that of an ASE source is in the THz regime. We thus conclude that the ASE source would outperform a coherent state probe because of its larger bandwidth, under identical thermal environments.

\emph{Performance evaluation.}~Let us consider sensing over a mono-static free-space channel of line-of-sight range $L$, with an ASE source of center wavelength $\lambda$. We employ a single spatial mode (the fundamental Gaussian mode) probe. Assuming vacuum propagation (ignoring turbulence and atmospheric extinction), and Gaussian-attenuation apertures (for simplifying the expressions without losing the essence of the problem), we get the transmissivities of both the forward and return paths as $\eta_1=\eta_2=\eta = (A_t A_T)/ (\lambda L)^2$, where $A_t = \pi r_t^2$ is the area of the radius-$r_t$ exit pupil of Alice's transmitter aperture, which we take to be equal to the area of the entrance pupil of the aperture of her receiver telescope. $A_T = \pi r_T^2$ is the effective (radius-$r_T$ circular) area of the target cross-section interrogated. The cw ASE source of optical bandwidth $W$ illuminates the target for $T$ seconds, which results in $n \approx WT$ temporal modes to probe the target.

\begin{figure}[t]
	\centering
	\includegraphics[width=1\linewidth]{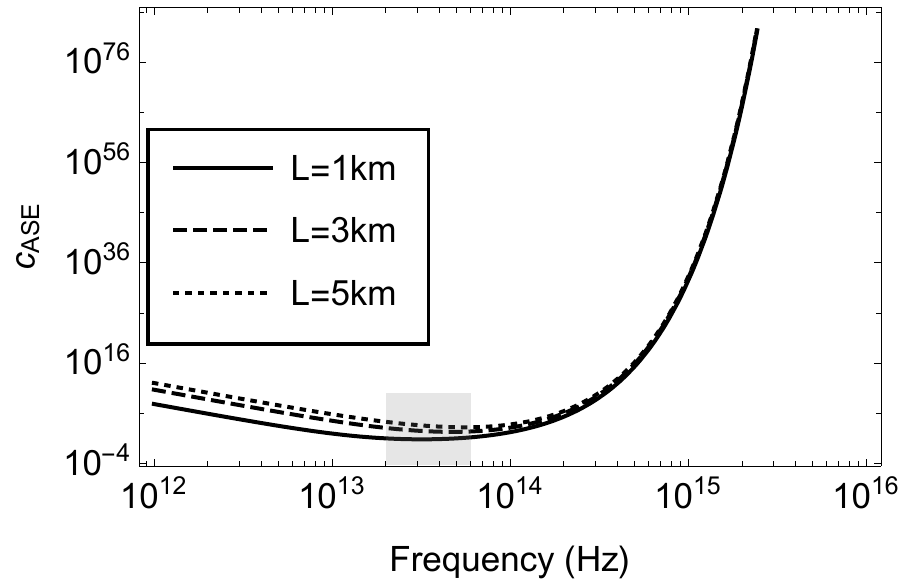}
	\caption{Plot of $c_{\textrm{ASE}}$ as a function of center frequency of transmission $f$, for three different values of target range, $L$. We used $r_t=4$ cm, $r_T= 10$ cm, and $\epsilon=10^{-3}$. Shaded area is where the minima appear.}
	\label{fig:cASEvFreq}
\end{figure}

In Fig. \ref{fig:cASEvFreq}, we plot $c_{\textrm{ASE}}$ of Eq. (\ref{MSEASE}) as a function of the center frequency $f = hc/\lambda$ of the source, where $h$ is Planck constant and $c$ is the speed of light. We assume Planck law limited thermal noise, with ${\bar n}_{B_1} = {\bar n}_{B_2} = {\bar n}_{B} = [\exp(hc/(\lambda k_B T_0))-1]^{-1}$, where $k_B$ is the Boltzmann constant and $T_0 = 300$K is the ambient temperature. Longer wavelengths (smaller frequencies) have more noise (larger ${\bar n}_B$) to hide the probe, which tends to make $c_{\rm ASE}$ smaller. On the other hand, longer wavelengths (smaller frequencies) have more diffraction limited loss (smaller $\eta$), giving more photons to Willie, which tends to make $c_{\rm ASE}$ higher. So, there is a trade-off in the choice of the center frequency, and we see that the minimum of $c_{\rm ASE}$ happens in the mid to long wave infrared (ML-IR) region. We numerically evaluate the optimal wavelength $\lambda$ which minimizes $c_{\textrm{ASE}}$. For $\epsilon=10^{-3}$, probing time $T=1$s, and for a source with bandwidth $W=3\textrm{THz}$ we find the right hand side of Eq. (\ref{MSEASE}), which we denote as $\textrm{B}$, the MSE lower bound. For $L=1\textrm{km}$: $\lambda=9.40 \mu\textrm{m}$, $\textrm{B}=0.00322$, for $L=3\textrm{km}$: $\lambda=6.35\mu\textrm{m}$, $\textrm{B}=0.09927$, and for $L=5\textrm{km}$: $\lambda=5.38\mu\textrm{m}$, $\textrm{B}=0.81438$.       

Sources in the ML-IR region with $W=3\textrm{THz}$ and wavelength $3\mu \textrm{m} \leq \lambda \leq 4.4 \mu \textrm{m}$ and $\lambda=4,4.6,8.7,10.2\mu \textrm{m}$ such as \cite{Neely2011}\cite{[{Personal communication with Dr. Bob Shine, DRS Daylight solutions, concerning the Aries-2 laser specifications}],Aries2}, for $L=1\textrm{km}$ and $T=1s$ give $\textrm{B}=0.00327$ and $\textrm{B}=0.08423$ for wavelengths $8.7\mu\textrm{m}$ and $3\mu\textrm{m}$ respectively, which correspond to the extrema of the performance for the aforementioned parameters, wavelength range and values.

\emph{Conclusions.}~Our work introduces to the greater communities of optical sensing, strategic security, and fundamental quantum optics, the concept of quantum-secure covert (low probability of detect) sensing. Our analytical analysis has shown that using floodlight illumination from an ASE source, covert phase sensing is possible and experimentally feasible. In fact, the ASE source enhances covert sensing performance by exploiting its large optical bandwidth $W$, i.e., the number of available orthogonal modes $n=WT$ can significantly increase for a given time $T$. An open question is to consider sensing multiple parameters, e.g., covert sensing of phase and loss (ranging) or multiple phases that could be the pixels of an image. We leave these questions for future work.
Our setup is experimentally feasible. The ASE source is divided into two arms by a highly reflective fiber coupler, so that  $\bar{n}_{\textrm{S}} \ll 1$ and $\bar{n}_{\textrm{LO}} \gg 1$. The sensing arm is frequency shifted prior to being sent to a very noisy environment and then is phase shifted by a weakly reflecting object. The LO is stored locally in a spool of optical fibers for later heterodyne measurement. Note that for sources with smaller wavelengths than those in \cite{Neely2011,[{Personal communication with Dr. Bob Shine, DRS Daylight solutions, concerning the Aries-2 laser specifications}] Aries2}, one could inject artificial noise to get a reasonable MSE lower bound for a proof of concept experiment. Our present work indicates covert sensing is experimentally feasible and applicable to practical optical sensing tasks. A thorough analysis of modeling, experimental demonstration, and application development, are left for follow up works.\\

\emph{Acknowledgments.}~CG was supported on a General Dynamics research and development contract to the University of Arizona. SG, BB and ZZ acknowledge the Office of Naval Research program Communications and Networking with Quantum Operationally-Secure Technology for Maritime Deployment (CONQUEST), awarded under Raytheon BBN Technologies prime contract number N00014-16-C-2069, and subcontract to University of Arizona. AD was partly supported by the UK EPSRC (EP/K04057X/2), and the UK National Quantum Technologies Programme (EP/M013243/1, EP/M01326X/1). The authors thank Darlene Hart (General Dynamics), William Clark (General Dynamics), and Bob Shine (DRS Daylight Solutions) for useful discussions on light sources.

\bibliography{covertFLbib,covertFLbibSM}

\begin{thebibliography}{26}%
\makeatletter
\providecommand \@ifxundefined [1]{%
 \@ifx{#1\undefined}
}%
\providecommand \@ifnum [1]{%
 \ifnum #1\expandafter \@firstoftwo
 \else \expandafter \@secondoftwo
 \fi
}%
\providecommand \@ifx [1]{%
 \ifx #1\expandafter \@firstoftwo
 \else \expandafter \@secondoftwo
 \fi
}%
\providecommand \natexlab [1]{#1}%
\providecommand \enquote  [1]{``#1''}%
\providecommand \bibnamefont  [1]{#1}%
\providecommand \bibfnamefont [1]{#1}%
\providecommand \citenamefont [1]{#1}%
\providecommand \href@noop [0]{\@secondoftwo}%
\providecommand \href [0]{\begingroup \@sanitize@url \@href}%
\providecommand \@href[1]{\@@startlink{#1}\@@href}%
\providecommand \@@href[1]{\endgroup#1\@@endlink}%
\providecommand \@sanitize@url [0]{\catcode `\\12\catcode `\$12\catcode
  `\&12\catcode `\#12\catcode `\^12\catcode `\_12\catcode `\%12\relax}%
\providecommand \@@startlink[1]{}%
\providecommand \@@endlink[0]{}%
\providecommand \url  [0]{\begingroup\@sanitize@url \@url }%
\providecommand \@url [1]{\endgroup\@href {#1}{\urlprefix }}%
\providecommand \urlprefix  [0]{URL }%
\providecommand \Eprint [0]{\href }%
\providecommand \doibase [0]{http://dx.doi.org/}%
\providecommand \selectlanguage [0]{\@gobble}%
\providecommand \bibinfo  [0]{\@secondoftwo}%
\providecommand \bibfield  [0]{\@secondoftwo}%
\providecommand \translation [1]{[#1]}%
\providecommand \BibitemOpen [0]{}%
\providecommand \bibitemStop [0]{}%
\providecommand \bibitemNoStop [0]{.\EOS\space}%
\providecommand \EOS [0]{\spacefactor3000\relax}%
\providecommand \BibitemShut  [1]{\csname bibitem#1\endcsname}%
\let\auto@bib@innerbib\@empty
\bibitem [{\citenamefont {Giovannetti}\ \emph {et~al.}(2006)\citenamefont
  {Giovannetti}, \citenamefont {Lloyd},\ and\ \citenamefont
  {Maccone}}]{Giov2006}%
  \BibitemOpen
  \bibfield  {author} {\bibinfo {author} {\bibfnamefont {V.}~\bibnamefont
  {Giovannetti}}, \bibinfo {author} {\bibfnamefont {S.}~\bibnamefont {Lloyd}},
  \ and\ \bibinfo {author} {\bibfnamefont {L.}~\bibnamefont {Maccone}},\ }\href
  {\doibase 10.1103/PhysRevLett.96.010401} {\bibfield  {journal} {\bibinfo
  {journal} {Phys. Rev. Lett.}\ }\textbf {\bibinfo {volume} {96}},\ \bibinfo
  {pages} {010401} (\bibinfo {year} {2006})}\BibitemShut {NoStop}%
\bibitem [{\citenamefont {Ono}\ \emph {et~al.}(2013)\citenamefont {Ono},
  \citenamefont {Okamoto},\ and\ \citenamefont {Takeuchi}}]{micro1}%
  \BibitemOpen
  \bibfield  {author} {\bibinfo {author} {\bibfnamefont {T.}~\bibnamefont
  {Ono}}, \bibinfo {author} {\bibfnamefont {R.}~\bibnamefont {Okamoto}}, \ and\
  \bibinfo {author} {\bibfnamefont {S.}~\bibnamefont {Takeuchi}},\ }\href
  {\doibase 10.1038/ncomms3426} {\bibfield  {journal} {\bibinfo  {journal}
  {Nature Communications}\ }\textbf {\bibinfo {volume} {4}} (\bibinfo {year}
  {2013}),\ 10.1038/ncomms3426}\BibitemShut {NoStop}%
\bibitem [{\citenamefont {Israel}\ \emph {et~al.}(2014)\citenamefont {Israel},
  \citenamefont {Rosen},\ and\ \citenamefont {Silberberg}}]{micro2}%
  \BibitemOpen
  \bibfield  {author} {\bibinfo {author} {\bibfnamefont {Y.}~\bibnamefont
  {Israel}}, \bibinfo {author} {\bibfnamefont {S.}~\bibnamefont {Rosen}}, \
  and\ \bibinfo {author} {\bibfnamefont {Y.}~\bibnamefont {Silberberg}},\
  }\href {\doibase 10.1103/PhysRevLett.112.103604} {\bibfield  {journal}
  {\bibinfo  {journal} {Phys. Rev. Lett.}\ }\textbf {\bibinfo {volume} {112}},\
  \bibinfo {pages} {103604} (\bibinfo {year} {2014})}\BibitemShut {NoStop}%
\bibitem [{\citenamefont {Rembe}\ \emph {et~al.}(2016)\citenamefont {Rembe},
  \citenamefont {Kadner},\ and\ \citenamefont {Giesen}}]{Rembe2017}%
  \BibitemOpen
  \bibfield  {author} {\bibinfo {author} {\bibfnamefont {C.}~\bibnamefont
  {Rembe}}, \bibinfo {author} {\bibfnamefont {L.}~\bibnamefont {Kadner}}, \
  and\ \bibinfo {author} {\bibfnamefont {M.}~\bibnamefont {Giesen}},\ }\href
  {\doibase 10.1063/1.4964625} {\bibfield  {journal} {\bibinfo  {journal}
  {Review of Scientific Instruments}\ }\textbf {\bibinfo {volume} {87}},\
  \bibinfo {pages} {102503} (\bibinfo {year} {2016})},\ \Eprint
  {http://arxiv.org/abs/https://doi.org/10.1063/1.4964625}
  {https://doi.org/10.1063/1.4964625} \BibitemShut {NoStop}%
\bibitem [{\citenamefont {Rarity}\ \emph {et~al.}(1990)\citenamefont {Rarity},
  \citenamefont {Tapster}, \citenamefont {Walker},\ and\ \citenamefont
  {Seward}}]{Rarity1990}%
  \BibitemOpen
  \bibfield  {author} {\bibinfo {author} {\bibfnamefont {J.~G.}\ \bibnamefont
  {Rarity}}, \bibinfo {author} {\bibfnamefont {P.~R.}\ \bibnamefont {Tapster}},
  \bibinfo {author} {\bibfnamefont {J.~G.}\ \bibnamefont {Walker}}, \ and\
  \bibinfo {author} {\bibfnamefont {S.}~\bibnamefont {Seward}},\ }\href
  {\doibase 10.1364/AO.29.002939} {\bibfield  {journal} {\bibinfo  {journal}
  {Appl. Opt.}\ }\textbf {\bibinfo {volume} {29}},\ \bibinfo {pages} {2939}
  (\bibinfo {year} {1990})}\BibitemShut {NoStop}%
\bibitem [{\citenamefont {Tsang}(2017)}]{Mankei2017}%
  \BibitemOpen
  \bibfield  {author} {\bibinfo {author} {\bibfnamefont {M.}~\bibnamefont
  {Tsang}},\ }\href {http://stacks.iop.org/1367-2630/19/i=2/a=023054}
  {\bibfield  {journal} {\bibinfo  {journal} {New Journal of Physics}\ }\textbf
  {\bibinfo {volume} {19}},\ \bibinfo {pages} {023054} (\bibinfo {year}
  {2017})}\BibitemShut {NoStop}%
\bibitem [{\citenamefont {Dale}\ and\ \citenamefont
  {Morley}(2017)}]{Gavin2017}%
  \BibitemOpen
  \bibfield  {author} {\bibinfo {author} {\bibfnamefont {M.~W.}\ \bibnamefont
  {Dale}}\ and\ \bibinfo {author} {\bibfnamefont {G.~W.}\ \bibnamefont
  {Morley}},\ }\href {https://arxiv.org/abs/1705.01994} {\bibfield  {journal}
  {\bibinfo  {journal} {arXiv preprint arXiv:1705.01994}\ } (\bibinfo {year}
  {2017})}\BibitemShut {NoStop}%
\bibitem [{\citenamefont {Abbott}\ \emph {et~al.}(2016)\citenamefont {Abbott}
  \emph {et~al.}}]{GW1}%
  \BibitemOpen
  \bibfield  {author} {\bibinfo {author} {\bibfnamefont {B.~P.}\ \bibnamefont
  {Abbott}} \emph {et~al.} (\bibinfo {collaboration} {LIGO Scientific
  Collaboration and Virgo Collaboration}),\ }\href {\doibase
  10.1103/PhysRevLett.116.061102} {\bibfield  {journal} {\bibinfo  {journal}
  {Phys. Rev. Lett.}\ }\textbf {\bibinfo {volume} {116}},\ \bibinfo {pages}
  {061102} (\bibinfo {year} {2016})}\BibitemShut {NoStop}%
\bibitem [{\citenamefont {Demkowicz-Dobrza\'{n}ski}\ \emph
  {et~al.}(2015)\citenamefont {Demkowicz-Dobrza\'{n}ski}, \citenamefont
  {Jarzyna},\ and\ \citenamefont
  {Ko\l{}ody\'{n}ski}}]{demkowicz15quantmetrologysurvey}%
  \BibitemOpen
  \bibfield  {author} {\bibinfo {author} {\bibfnamefont {R.}~\bibnamefont
  {Demkowicz-Dobrza\'{n}ski}}, \bibinfo {author} {\bibfnamefont
  {M.}~\bibnamefont {Jarzyna}}, \ and\ \bibinfo {author} {\bibfnamefont
  {J.}~\bibnamefont {Ko\l{}ody\'{n}ski}},\ }\href {\doibase
  10.1016/bs.po.2015.02.003} {\bibfield  {journal} {\bibinfo  {journal}
  {Progress in Optics}\ }\textbf {\bibinfo {volume} {60}},\ \bibinfo {pages}
  {345} (\bibinfo {year} {2015})},\ \Eprint
  {http://arxiv.org/abs/arXiv:1405.7703} {arXiv:1405.7703} \BibitemShut
  {NoStop}%
\bibitem [{\citenamefont {Bash}\ \emph {et~al.}(2012)\citenamefont {Bash},
  \citenamefont {Goeckel},\ and\ \citenamefont {Towsley}}]{Bash2012}%
  \BibitemOpen
  \bibfield  {author} {\bibinfo {author} {\bibfnamefont {B.~A.}\ \bibnamefont
  {Bash}}, \bibinfo {author} {\bibfnamefont {D.}~\bibnamefont {Goeckel}}, \
  and\ \bibinfo {author} {\bibfnamefont {D.}~\bibnamefont {Towsley}},\
  }\href@noop {} {\emph {\bibinfo {title} {Limits of Reliable Communication
  with Low Probability of Detection on {AWGN} Channels}}},\ \bibinfo {type}
  {Tech. Rep.}\ \bibinfo {number} {UM-CS-2012-003}\ (\bibinfo  {institution}
  {University of Massachusetts},\ \bibinfo {year} {2012})\BibitemShut {NoStop}%
\bibitem [{\citenamefont {Che}\ \emph {et~al.}(2013)\citenamefont {Che},
  \citenamefont {Bakshi},\ and\ \citenamefont {Jaggi}}]{che13sqrtlawbsc}%
  \BibitemOpen
  \bibfield  {author} {\bibinfo {author} {\bibfnamefont {P.~H.}\ \bibnamefont
  {Che}}, \bibinfo {author} {\bibfnamefont {M.}~\bibnamefont {Bakshi}}, \ and\
  \bibinfo {author} {\bibfnamefont {S.}~\bibnamefont {Jaggi}},\ }\href@noop {}
  {\enquote {\bibinfo {title} {Reliable deniable communication: Hiding messages
  in noise},}\ }\bibinfo {howpublished} {arXiv:1304.6693} (\bibinfo {year}
  {2013})\BibitemShut {NoStop}%
\bibitem [{\citenamefont {Wang}\ \emph {et~al.}(2016)\citenamefont {Wang},
  \citenamefont {Wornell},\ and\ \citenamefont {Zheng}}]{wang15covert}%
  \BibitemOpen
  \bibfield  {author} {\bibinfo {author} {\bibfnamefont {L.}~\bibnamefont
  {Wang}}, \bibinfo {author} {\bibfnamefont {G.~W.}\ \bibnamefont {Wornell}}, \
  and\ \bibinfo {author} {\bibfnamefont {L.}~\bibnamefont {Zheng}},\ }\href
  {\doibase 10.1109/TIT.2016.2548471} {\bibfield  {journal} {\bibinfo
  {journal} {IEEE Trans. Inf. Theory}\ }\textbf {\bibinfo {volume} {62}},\
  \bibinfo {pages} {3493} (\bibinfo {year} {2016})}\BibitemShut {NoStop}%
\bibitem [{\citenamefont {Bash}\ \emph {et~al.}(2015)\citenamefont {Bash},
  \citenamefont {Gheorghe}, \citenamefont {Patel}, \citenamefont {Habif},
  \citenamefont {Goeckel}, \citenamefont {Towsley},\ and\ \citenamefont
  {Guha}}]{bash15covertbosoniccomm}%
  \BibitemOpen
  \bibfield  {author} {\bibinfo {author} {\bibfnamefont {B.~A.}\ \bibnamefont
  {Bash}}, \bibinfo {author} {\bibfnamefont {A.~H.}\ \bibnamefont {Gheorghe}},
  \bibinfo {author} {\bibfnamefont {M.}~\bibnamefont {Patel}}, \bibinfo
  {author} {\bibfnamefont {J.~L.}\ \bibnamefont {Habif}}, \bibinfo {author}
  {\bibfnamefont {D.}~\bibnamefont {Goeckel}}, \bibinfo {author} {\bibfnamefont
  {D.}~\bibnamefont {Towsley}}, \ and\ \bibinfo {author} {\bibfnamefont
  {S.}~\bibnamefont {Guha}},\ }\href {\doibase 10.1038/NCOMMS9626} {\bibfield
  {journal} {\bibinfo  {journal} {Nat Commun}\ }\textbf {\bibinfo {volume} {6}}
  (\bibinfo {year} {2015}),\ 10.1038/NCOMMS9626}\BibitemShut {NoStop}%
\bibitem [{\citenamefont {Bash}\ and\ \citenamefont
  {Guha}(2018)}]{PatentBash2018}%
  \BibitemOpen
  \bibfield  {author} {\bibinfo {author} {\bibfnamefont {B.~A.}\ \bibnamefont
  {Bash}}\ and\ \bibinfo {author} {\bibfnamefont {S.}~\bibnamefont {Guha}},\
  }\href {http://www.freepatentsonline.com/y2018/0210071.html} {\enquote
  {\bibinfo {title} {Covert sensor},}\ } (\bibinfo {year} {2018}),\ \bibinfo
  {note} {\ United States Patent Application number 15/875,904}\BibitemShut
  {NoStop}%
\bibitem [{\citenamefont {Bash}\ \emph {et~al.}(2017)\citenamefont {Bash},
  \citenamefont {Gagatsos}, \citenamefont {Datta},\ and\ \citenamefont
  {Guha}}]{bash2017isit}%
  \BibitemOpen
  \bibfield  {author} {\bibinfo {author} {\bibfnamefont {B.~A.}\ \bibnamefont
  {Bash}}, \bibinfo {author} {\bibfnamefont {C.~N.}\ \bibnamefont {Gagatsos}},
  \bibinfo {author} {\bibfnamefont {A.}~\bibnamefont {Datta}}, \ and\ \bibinfo
  {author} {\bibfnamefont {S.}~\bibnamefont {Guha}},\ }in\ \href {\doibase
  10.1109/ISIT.2017.8007122} {\emph {\bibinfo {booktitle} {2017 IEEE
  International Symposium on Information Theory (ISIT)}}}\ (\bibinfo {year}
  {2017})\ pp.\ \bibinfo {pages} {3210--3214},\ \Eprint
  {http://arxiv.org/abs/arXiv:1701.06206v1} {arXiv:1701.06206v1} \BibitemShut
  {NoStop}%
\bibitem [{\citenamefont {Zhuang}\ \emph {et~al.}(2016)\citenamefont {Zhuang},
  \citenamefont {Zhang}, \citenamefont {Dove}, \citenamefont {Wong},\ and\
  \citenamefont {Shapiro}}]{zhuang15floodlightQKD}%
  \BibitemOpen
  \bibfield  {author} {\bibinfo {author} {\bibfnamefont {Q.}~\bibnamefont
  {Zhuang}}, \bibinfo {author} {\bibfnamefont {Z.}~\bibnamefont {Zhang}},
  \bibinfo {author} {\bibfnamefont {J.}~\bibnamefont {Dove}}, \bibinfo {author}
  {\bibfnamefont {F.~N.~C.}\ \bibnamefont {Wong}}, \ and\ \bibinfo {author}
  {\bibfnamefont {J.~H.}\ \bibnamefont {Shapiro}},\ }\href {\doibase
  10.1103/PhysRevA.94.012322} {\bibfield  {journal} {\bibinfo  {journal} {Phys.
  Rev. A}\ }\textbf {\bibinfo {volume} {94}},\ \bibinfo {pages} {012322}
  (\bibinfo {year} {2016})},\ \bibinfo {note} {arXiv:1510.08737
  [quant-ph]}\BibitemShut {NoStop}%
\bibitem [{\citenamefont {Escher}\ \emph {et~al.}(2011)\citenamefont {Escher},
  \citenamefont {Filho},\ and\ \citenamefont {Davidovich}}]{Escher2011}%
  \BibitemOpen
  \bibfield  {author} {\bibinfo {author} {\bibfnamefont {R.~M.}\ \bibnamefont
  {Escher}}, \bibinfo {author} {\bibfnamefont {R.~L.~M.}\ \bibnamefont
  {Filho}}, \ and\ \bibinfo {author} {\bibfnamefont {L.}~\bibnamefont
  {Davidovich}},\ }\href {\doibase 10.1038/NPHYS1958} {\bibfield  {journal}
  {\bibinfo  {journal} {Nature Physics}\ }\textbf {\bibinfo {volume} {7}},\
  \bibinfo {pages} {406} (\bibinfo {year} {2011})}\BibitemShut {NoStop}%
\bibitem [{\citenamefont {Gagatsos}\ \emph {et~al.}(2017)\citenamefont
  {Gagatsos}, \citenamefont {Bash}, \citenamefont {Guha},\ and\ \citenamefont
  {Datta}}]{Gagatsos2017bound}%
  \BibitemOpen
  \bibfield  {author} {\bibinfo {author} {\bibfnamefont {C.~N.}\ \bibnamefont
  {Gagatsos}}, \bibinfo {author} {\bibfnamefont {B.~A.}\ \bibnamefont {Bash}},
  \bibinfo {author} {\bibfnamefont {S.}~\bibnamefont {Guha}}, \ and\ \bibinfo
  {author} {\bibfnamefont {A.}~\bibnamefont {Datta}},\ }\href {\doibase
  10.1103/PhysRevA.96.062306} {\bibfield  {journal} {\bibinfo  {journal} {Phys.
  Rev. A}\ }\textbf {\bibinfo {volume} {96}},\ \bibinfo {pages} {062306}
  (\bibinfo {year} {2017})}\BibitemShut {NoStop}%
\bibitem [{\citenamefont {Neely}\ \emph {et~al.}(2011)\citenamefont {Neely},
  \citenamefont {Johnson},\ and\ \citenamefont {Diddams}}]{Neely2011}%
  \BibitemOpen
  \bibfield  {author} {\bibinfo {author} {\bibfnamefont {T.~W.}\ \bibnamefont
  {Neely}}, \bibinfo {author} {\bibfnamefont {T.~A.}\ \bibnamefont {Johnson}},
  \ and\ \bibinfo {author} {\bibfnamefont {S.~A.}\ \bibnamefont {Diddams}},\
  }\href {\doibase 10.1364/OL.36.004020} {\bibfield  {journal} {\bibinfo
  {journal} {Opt. Lett.}\ }\textbf {\bibinfo {volume} {36}},\ \bibinfo {pages}
  {4020} (\bibinfo {year} {2011})}\BibitemShut {NoStop}%
\bibitem [{Ari()}]{Aries2}%
  \BibitemOpen
  \href@noop {} {}\BibitemShut {NoStop}%
\bibitem [{\citenamefont {Weedbrook}\ \emph {et~al.}(2012)\citenamefont
  {Weedbrook}, \citenamefont {Pirandola}, \citenamefont {Garc\'{\i}a-Patr\'on},
  \citenamefont {Cerf}, \citenamefont {Ralph}, \citenamefont {Shapiro},\ and\
  \citenamefont {Lloyd}}]{Review2012}%
  \BibitemOpen
  \bibfield  {author} {\bibinfo {author} {\bibfnamefont {C.}~\bibnamefont
  {Weedbrook}}, \bibinfo {author} {\bibfnamefont {S.}~\bibnamefont
  {Pirandola}}, \bibinfo {author} {\bibfnamefont {R.}~\bibnamefont
  {Garc\'{\i}a-Patr\'on}}, \bibinfo {author} {\bibfnamefont {N.~J.}\
  \bibnamefont {Cerf}}, \bibinfo {author} {\bibfnamefont {T.~C.}\ \bibnamefont
  {Ralph}}, \bibinfo {author} {\bibfnamefont {J.~H.}\ \bibnamefont {Shapiro}},
  \ and\ \bibinfo {author} {\bibfnamefont {S.}~\bibnamefont {Lloyd}},\ }\href
  {\doibase 10.1103/RevModPhys.84.621} {\bibfield  {journal} {\bibinfo
  {journal} {Rev. Mod. Phys.}\ }\textbf {\bibinfo {volume} {84}},\ \bibinfo
  {pages} {621} (\bibinfo {year} {2012})}\BibitemShut {NoStop}%
\bibitem [{\citenamefont {Pirandola}\ \emph {et~al.}(2017)\citenamefont
  {Pirandola}, \citenamefont {Laurenza}, \citenamefont {Ottaviani},\ and\
  \citenamefont {Banchi}}]{Pirandola2017}%
  \BibitemOpen
  \bibfield  {author} {\bibinfo {author} {\bibfnamefont {S.}~\bibnamefont
  {Pirandola}}, \bibinfo {author} {\bibfnamefont {R.}~\bibnamefont {Laurenza}},
  \bibinfo {author} {\bibfnamefont {C.}~\bibnamefont {Ottaviani}}, \ and\
  \bibinfo {author} {\bibfnamefont {L.}~\bibnamefont {Banchi}},\ }\href
  {https://doi.org/10.1038/ncomms15043} {\bibfield  {journal} {\bibinfo
  {journal} {Nature Communications}\ }\textbf {\bibinfo {volume} {8}},\
  \bibinfo {pages} {15043} (\bibinfo {year} {2017})}\BibitemShut {NoStop}%
\bibitem [{\citenamefont {Topsoe}(2004)}]{LogBounds}%
  \BibitemOpen
  \bibfield  {author} {\bibinfo {author} {\bibfnamefont {F.}~\bibnamefont
  {Topsoe}},\ }\href {https://rgmia.org/v7n2.php} {\bibfield  {journal}
  {\bibinfo  {journal} {Inequality Theory and Applications}\ }\textbf {\bibinfo
  {volume} {7}} (\bibinfo {year} {2004})}\BibitemShut {NoStop}%
\bibitem [{\citenamefont {Caruso}\ \emph {et~al.}(2008)\citenamefont {Caruso},
  \citenamefont {Eisert}, \citenamefont {Giovannetti},\ and\ \citenamefont
  {Holevo}}]{Holevo2008}%
  \BibitemOpen
  \bibfield  {author} {\bibinfo {author} {\bibfnamefont {F.}~\bibnamefont
  {Caruso}}, \bibinfo {author} {\bibfnamefont {J.}~\bibnamefont {Eisert}},
  \bibinfo {author} {\bibfnamefont {V.}~\bibnamefont {Giovannetti}}, \ and\
  \bibinfo {author} {\bibfnamefont {A.~S.}\ \bibnamefont {Holevo}},\ }\href
  {http://stacks.iop.org/1367-2630/10/i=8/a=083030} {\bibfield  {journal}
  {\bibinfo  {journal} {New Journal of Physics}\ }\textbf {\bibinfo {volume}
  {10}},\ \bibinfo {pages} {083030} (\bibinfo {year} {2008})}\BibitemShut
  {NoStop}%
\bibitem [{\citenamefont {Marian}\ and\ \citenamefont
  {Marian}(2012)}]{Marian2012}%
  \BibitemOpen
  \bibfield  {author} {\bibinfo {author} {\bibfnamefont {P.}~\bibnamefont
  {Marian}}\ and\ \bibinfo {author} {\bibfnamefont {T.~A.}\ \bibnamefont
  {Marian}},\ }\href {\doibase 10.1103/PhysRevA.86.022340} {\bibfield
  {journal} {\bibinfo  {journal} {Phys. Rev. A}\ }\textbf {\bibinfo {volume}
  {86}},\ \bibinfo {pages} {022340} (\bibinfo {year} {2012})}\BibitemShut
  {NoStop}%
\bibitem [{\citenamefont {Banchi}\ \emph {et~al.}(2015)\citenamefont {Banchi},
  \citenamefont {Braunstein},\ and\ \citenamefont {Pirandola}}]{Banchi2015}%
  \BibitemOpen
  \bibfield  {author} {\bibinfo {author} {\bibfnamefont {L.}~\bibnamefont
  {Banchi}}, \bibinfo {author} {\bibfnamefont {S.~L.}\ \bibnamefont
  {Braunstein}}, \ and\ \bibinfo {author} {\bibfnamefont {S.}~\bibnamefont
  {Pirandola}},\ }\href {\doibase 10.1103/PhysRevLett.115.260501} {\bibfield
  {journal} {\bibinfo  {journal} {Phys. Rev. Lett.}\ }\textbf {\bibinfo
  {volume} {115}},\ \bibinfo {pages} {260501} (\bibinfo {year}
  {2015})}\BibitemShut {NoStop}%
\end{thebibliography}%

\newpage

\onecolumngrid
\appendix
\section*{Appendix}
\renewcommand{\thesubsection}{\arabic{subsection}}
\def\theequation{A\arabic{equation}}
\setcounter{equation}{0}

\noindent
We give information on the derivations used in the main paper. Specifically, we discuss detectability, we derive the covariance matrices for the global output, for Alice, and for Willie. We calculate the quantum relative entropy (QRE) for Willie's part between the states which correspond to \emph{Alice is sensing} and \emph{Alice is not sensing} scenarios and we discuss the Taylor expansion of the QRE. We prove that Alice sees only one effective thermal channel. We give explicit calculations for the heterodyne detection. Finally, we derive the quantum Fisher information for when Alice is sensing the unknown phase. 

\subsection{Detectability}\label{SM:Detect}
The adversary performs a binary hypothesis test on his sample to determine
whether the target is being interrogated or not.
Performance of the hypothesis test is typically measured by its
detection error probability denoted as $\mathbb{P}_{\rm e}^{\rm (det)}$ here since we speak generally, but it is $\mathbb{P}_{\rm e}^{\rm (w)}$ in the main paper when we talk about the ASE source. We have, 
$\mathbb{P}_{\rm e}^{\rm (det)}=\frac{\mathbb{P}_{\rm FA}+\mathbb{P}_{\rm MD}}{2}$,
where equal prior probabilities on sensor's interrogation state are
assumed, $\mathbb{P}_{\rm FA}$ is the probability of false alarm and
$\mathbb{P}_{\rm MD}$ is the probability of missed detection.
The sensor desires to remain covert by
ensuring that $\mathbb{P}_{\rm e}^{\rm (det)}\geq\frac{1}{2}-\epsilon$ for
an arbitrary small $\epsilon>0$ regardless of adversary's measurement
choice (since $\mathbb{P}_{\rm e}^{\rm (det)}=\frac{1}{2}$ for a random 
guess).
By decreasing the power used in a probe, the sensor can decrease
the effectiveness of the adversary's hypothesis test at the expense of
the increased mean squared error (MSE) of the estimate.

\subsection{Alice's and Willie's covariance matrices}\label{SM:CM}
We give the full expressions of: the final covariance matrix (CM) of the global state, i.e, the final state that contains both Alice's and Willie's modes ($4$ modes in total, therefore the dimensions of the CM is $8\times 8$), and the CM of each one Alice and Willie. We remind the reader that the first moments are zero. We also provide a guide on how these CM were derived.

We use the symplectic formalism \cite{Review2012}, where under a symplectic matrix $S$, a CM $V_0$ is evolved to the CM $V=SV_0S^T$. All CM are in the $qqpp$ representation, meaning that for a $2N \times 2N$ CM the upper left block contains information only on position, the lower right block concerns only momentum, and the off-diagonal blocks contain information on position-momentum correlations. The phase space vector is $r=\left(r_1,\ldots r_{2N}\right)^T=\left(q_1,\ldots q_N, p_1 \dots p_N \right)^T$. In our set up, we assume that the first two modes belong to Willie, while the last two modes belong to Alice. We have worked with $\hbar=1$, therefore the CM for vacuum is $I/2$, where $I$ is the identity matrix.     

Since initially Alice's and Willie's systems are uncorrelated, in the $qqpp$ representation the initial, global CM is
\begin{eqnarray}
	V_{{AW}_0}=
	\begin{pmatrix}
		\bar{n}_{\textrm{B}_2}+\frac{1}{2} & 0 & 0 & 0 & 0 & 0 & 0 & 0\\
		0 & \bar{n}_{\textrm{B}_1}+\frac{1}{2} & 0 & 0 & 0 & 0 & 0 & 0\\
		0 & 0 & \bar{n}_S+\frac{1}{2} & \sqrt{\bar{n}_\textrm{S} \bar{n}_{\textrm{LO}}} & 0 & 0 & 0 & 0\\
		0 & 0 & \sqrt{\bar{n}_\textrm{S} \bar{n}_{\textrm{LO}}} & \bar{n}_\textrm{LO} + \frac{1}{2} & 0 & 0 & 0 & 0\\
		0 & 0 & 0 & 0 & \bar{n}_{\textrm{B}_2}+\frac{1}{2} & 0 & 0 & 0\\
		0 & 0 & 0 & 0 & 0 & \bar{n}_{\textrm{B}_1}+\frac{1}{2} & 0 & 0\\
		0 & 0 & 0 & 0 & 0 & 0 & \bar{n}_S+\frac{1}{2} & \sqrt{\bar{n}_\textrm{S} \bar{n}_{\textrm{LO}}}\\
		0 & 0 & 0 & 0 & 0 & 0 & \sqrt{\bar{n}_\textrm{S} \bar{n}_{\textrm{LO}}} & \bar{n}_\textrm{LO} + \frac{1}{2}
	\end{pmatrix}
\end{eqnarray}

\begin{figure}[h]
	\centering
	\includegraphics[width=0.8\linewidth]{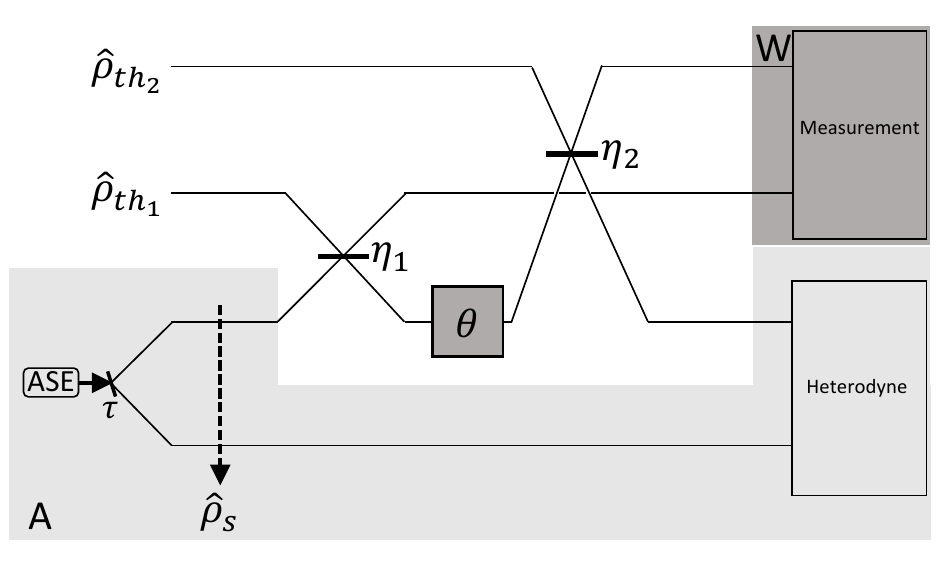}
	\caption{A layout of the setup, equivalent to the one of the main paper, however more explanatory regarding the transformations. Light gray area (A) belongs to Alice, while darker gray area (W) belongs to Willie.}
	\label{fig:SMthermal2}
\end{figure}

Our setup consists of three symplectic transformations, which are depicted in Fig.~\ref{fig:SMthermal2} and in order of action are the following. A beam splitter $B_{23}(\eta_1)$ with transmissivity $\eta_1$ acting on modes $2$ and $3$, a phase shift $R_3(\theta)$ acting on mode $3$, and a beam splitter $B_{13}(\eta_2)$ with transmissivity $\eta_2$ acting on modes $1$ and $3$. By noticing on which modes each symplectic matrix acts, to find a the final CM becomes a straightforward job. Let us demonstrate an example. The beam splitter $B_{23}(\eta_1)$ will act on the sub-matrices $V^{(\textrm{pos}_1)}_{{AW}_0}$ and $V^{(\textrm{mom}_1)}_{{AW}_0}$ of $V_{{AW}_0}$ with elements $V_{{AW}_0}^{ij},\ i,j=2,3$ and $V_{{AW}_0}^{ij},\ i,j=6,7$ respectively, i.e., to the momentum and position coordinates with indices  $2,3$. The symplectic matrix $B_{23}(\eta_1)$ is,
\begin{eqnarray}
	B_{23}(\eta_1) =
	\begin{pmatrix}
		B_{23}^{(\textrm{pos})}(\eta_1) & 0_{2\times 2} \\
		0_{2\times 2} & B_{23}^{(\textrm{mom})}(\eta_1) 
	\end{pmatrix},
\end{eqnarray}
where
\begin{eqnarray}
	B_{23}^{\textrm{(pos)}}(\eta_1) = B_{23}^{\textrm{(mom)}}(\eta_1) =
	\begin{pmatrix}
		\sqrt{\eta_1} & \sqrt{1-\eta_1} \\
		-\sqrt{1-\eta_1} & \sqrt{\eta_1} 
	\end{pmatrix}.
\end{eqnarray}
Therefore we apply the transformation $V_1=B_{23}(\eta_1) \left( V^{(\textrm{pos}_1)}_{{AW}_0}\oplus V^{(\textrm{mom}_1)}_{{AW}_0}\right) B_{23}^T(\eta_1)$ where all involved matrices are block diagonal and therefore the resulting matrix will also be block-diagonal. Then in $V_{{AW}_0}$, we substitute the $V^{(\textrm{pos}_1)}_{{AW}_0}$ and $V^{(\textrm{mom}_1)}_{{AW}_0}$ blocks with the resulting blocks of $V_1$. The rest of $V_{{AW}_0}$ elements remain unchanged.

Subsequently, we apply the phase shift which corresponds to the symplectic matrix,
\begin{eqnarray}
	R_3(\theta)=
	\begin{pmatrix}
		1 & 0 & 0 & 0 & 0 & 0 & 0 & 0\\
		0 & 1 & 0 & 0 & 0 & 0 & 0 & 0\\
		0 & 0 & \cos\theta & 0 & 0 & 0 & -\sin\theta & 0\\
		0 & 0 & 0 & 1 & 0 & 0 & 0 & 0\\
		0 & 0 & 0 & 0 & 1 & 0 & 0 & 0\\
		0 & 0 & 0 & 0 & 0 & 1 & 0 & 0\\
		0 & 0 & \sin\theta & 0 & 0 & 0 & \cos\theta & 0\\
		0 & 0 & 0 & 0 & 0 & 0 & 0 & 1
	\end{pmatrix}
\end{eqnarray}
or one can again simplify calculations by noticing which modes the phase shift affects. Finally, the beam splitter $B_{13}(\eta_2)$ is applied on the resulting CM. The final, global CM is
\begin{eqnarray}
	\label{CMglobal}V=\begin{pmatrix}
		B & C \\
		C^T & B
	\end{pmatrix}
\end{eqnarray}
where,
\begin{eqnarray}
	B=
	\begin{pmatrix}
		B_W & B_{AW} \\
		B_{AW} & B_A
	\end{pmatrix}
\end{eqnarray}
and
\begin{eqnarray}
	C=
	\begin{pmatrix}
		C_W & C_{AW} \\
		-C_{AW}^T & C_A
	\end{pmatrix}
\end{eqnarray}
with
\begin{eqnarray}
	B_W=\begin{pmatrix}
		(1-\eta_2)\eta_1 \bar{n}_{\textrm{S}} + (1-\eta_1)(1-\eta_2)\bar{n}_{\textrm{B}_1}+\eta_2 \bar{n}_{\textrm{B}_2} + \frac{1}{2} & \sqrt{(1-\eta_2)\eta_1(1-\eta_1)} (\bar{n}_{\textrm{B}_1}-\bar{n}_{\textrm{S}}) \cos\theta \\
		\sqrt{(1-\eta_2)\eta_1(1-\eta_1)} (\bar{n}_{\textrm{B}_1}-\bar{n}_{\textrm{S}}) \cos\theta & \eta_1 \bar{n}_{\textrm{B}_1} + (1-\eta_1) \bar{n}_{\textrm{S}} + \frac{1}{2}
	\end{pmatrix},
	\label{BW}
\end{eqnarray}
\begin{eqnarray}
	B_{AW}=\begin{pmatrix}
		\sqrt{\eta_2(1-\eta_2)}[\eta_1 \bar{n}_{\textrm{S}}+(1-\eta_1)\bar{n}_{\textrm{B}_1}-\bar{n}_{\textrm{B}_2}] & \sqrt{(1-\eta_2)\eta_1 \bar{n}_{\textrm{S}} \bar{n}_{\textrm{LO}}} \cos\theta \\
		\sqrt{\eta_2 \eta_1 (1-\eta_1)}(\bar{n}_{\textrm{B}_1}-\bar{n}_{\textrm{S}})\cos\theta & \sqrt{(1-\eta_1)\bar{n}_{\textrm{S}} N_LO}
	\end{pmatrix},
\end{eqnarray}
\begin{eqnarray}
	B_A=\begin{pmatrix}
		(1-\eta_1)\eta_2 \bar{n}_{\textrm{B}_1} + (1-\eta_2)\bar{n}_{\textrm{B}_2}+\eta_1\eta_2 \bar{n}_{\textrm{S}} +\frac{1}{2} & \sqrt{\eta_1\eta_2 \bar{n}_{\textrm{S}} \bar{n}_{\textrm{LO}}} \cos\theta\\
		\sqrt{\eta_1\eta_2 \bar{n}_{\textrm{S}} \bar{n}_{\textrm{LO}}} \cos\theta & \bar{n}_{\textrm{LO}}+\frac{1}{2}
	\end{pmatrix},
\end{eqnarray}
\begin{eqnarray}
	C_W=\begin{pmatrix}
		0 & -\sqrt{(1-\eta_2)\eta_1(1-\eta_1)} (\bar{n}_{\textrm{B}_1}-\bar{n}_{\textrm{S}}) \sin\theta\\
		\sqrt{(1-\eta_2)\eta_1(1-\eta_1)} (\bar{n}_{\textrm{B}_1}-\bar{n}_{\textrm{S}}) \sin\theta & 0
	\end{pmatrix},
	\label{CW}
\end{eqnarray}
\begin{eqnarray}
	C_{AW}=\begin{pmatrix}
		0 & -\sqrt{(1-\eta_2)\eta_1 \bar{n}_{\textrm{S}} \bar{n}_{\textrm{LO}}}\sin\theta\\
		\sqrt{\eta_2\eta_1(1-\eta_1)} (\bar{n}_{\textrm{B}_1}-\bar{n}_{\textrm{S}}) \sin\theta & 0
	\end{pmatrix},
\end{eqnarray}
\begin{eqnarray}
	C_{A}=\begin{pmatrix}
		0 & -\sqrt{\eta_2\eta_1 \bar{n}_{\textrm{S}} \bar{n}_{\textrm{LO}}}\sin\theta\\
		\sqrt{\eta_2\eta_1 \bar{n}_{\textrm{S}} \bar{n}_{\textrm{LO}}}\sin\theta & 0
	\end{pmatrix}.
\end{eqnarray}

By tracing out, i.e, by keeping the corresponding sub-matrices, we can find the covariance matrices $V_A$ for Alice and $V_W$ for Willie,
\begin{eqnarray}
	\label{VA}V_A=
	\begin{pmatrix}
		B_A & C_{A}\\
		C_A^T & B_A
	\end{pmatrix}=
	\begin{pmatrix}
		a_{11} & -a_{12}\cos\theta & 0 & a_{12} \sin\theta \\
		-a_{12}\cos\theta & a_{22} & -a_{12} \sin\theta & 0 \\
		0 & -a_{12} \sin\theta & a_{11} & -a_{12}\cos\theta \\
		a_{12} \sin\theta & 0 & -a_{12} \cos\theta & a_{22}
	\end{pmatrix},
\end{eqnarray}
\begin{eqnarray}
	\label{VW}V_W=
	\begin{pmatrix}
		B_W & C_{W}\\
		C_W^T & B_W
	\end{pmatrix}=
	\begin{pmatrix}
		w_{11} & -w_{12}\cos\theta & 0 & w_{12} \sin\theta \\
		-w_{12}\cos\theta & w_{22} & -w_{12} \sin\theta & 0 \\
		0 & -w_{12} \sin\theta & w_{11} & -w_{12}\cos\theta \\
		w_{12} \sin\theta & 0 & -w_{12} \cos\theta & w_{22}
	\end{pmatrix},
\end{eqnarray}
where
\begin{eqnarray}
	\label{eq:a11} a_{11} &=& (1-\eta_1)\eta_2 \bar{n}_{\textrm{B}_1} + (1-\eta_2)\bar{n}_{\textrm{B}_2}+\eta_1\eta_2 \bar{n}_{\textrm{S}} +\frac{1}{2},\\
	\label{eq:a12} a_{12} &=& -\sqrt{\eta_1\eta_2 \bar{n}_{\textrm{S}} \bar{n}_{\textrm{LO}}} \cos\theta,\\
	\label{eq:a22} a_{22} &=& \bar{n}_{\textrm{LO}}+\frac{1}{2},\\
	w_{11} &=& (1-\eta_2)\eta_1 \bar{n}_\textrm{S} + (1-\eta_1)(1-\eta_2)\bar{n}_{\textrm{B}_1}+\eta_2 \bar{n}_{\textrm{B}_2} +\frac{1}{2},\\
	w_{12} &=& \sqrt{(1-\eta_2)\eta_1(1-\eta_1)} (\bar{n}_{\textrm{B}_1}-\bar{n}_\textrm{S}),\\
	w_{22} &=&  \eta_1 \bar{n}_{\textrm{B}_1} + (1-\eta_1) \bar{n}_\textrm{S} + \frac{1}{2}.
\end{eqnarray}

\subsection{The quantum relative entropy}\label{SM:QRE}
To calculate the quantum relative entropy (QRE) $D(\hat{\rho}_0\|\hat{\rho}_1)=\textrm{tr}\left(\hat{\rho}_0\ln \hat{\rho}_0 \right)-\textrm{tr}\left(\hat{\rho}_0\ln \hat{\rho}_1 \right)$ between two Gaussian states $\hat{\rho}_0$ and $\hat{\rho}_1$, we need to find the symplectic eigenvalues and symplectic eigenvectors of one of the the two covariance matrices and the symplectic eigenvalues of the other \cite{Pirandola2017} (section Methods therein). Let $\hat{\rho}_0$, $\hat{\rho}_1$ to have a covariance matrix $V_0$, $V_1$ respectively. The covariance matrix $V_1$ is for the case where Alice is sensing, i.e, ($\bar{n}_{\textrm{S}} \neq 0$), while the covariance matrix $V_0$ is for the case where Alice is not sensing, i.e., ($\bar{n}_{\textrm{S}} = 0$), therefore $V_0=V_W(\bar{n}_{\textrm{S}}=0)$ and $V_1=V_W(\bar{n}_{\textrm{S}} \neq 0)$, where $V_W$ is Willie's covariance matrix given in Eq. (\ref{VW}). That means Willie's CM matrix elements for $\bar{n}_{\textrm{S}}=0$, i.e, the matrix elements of $V_0$, are
\begin{eqnarray}
	\label{w011} w_{11}^{(0)} &=& (1-\eta_1)(1-\eta_2)\bar{n}_{\textrm{B}_1}+\eta_2 \bar{n}_{\textrm{B}_2} +\frac{1}{2},\\
	\label{w012} w_{12}^{(0)} &=& \sqrt{(1-\eta_2)\eta_1(1-\eta_1)} \bar{n}_{\textrm{B}_1},\\
	\label{w022} w_{22}^{(0)} &=&  \eta_1 \bar{n}_{\textrm{B}_1} + \frac{1}{2}
\end{eqnarray}
and $V_0$ will have the same structure as $V_W$ in Eq. (\ref{VW}). The QRE is,
\begin{eqnarray}
	\label{QRE1}D(\hat{\rho}_0\|\hat{\rho}_1) = 	-\Sigma(V_0,V_0)+\Sigma(V_0,V_1).
\end{eqnarray}
In Eq. (\ref{QRE1}), the first term is the von Neumann entropy (with a minus in front) of a Gaussian state with CM $V_0$ and zero first moments, which for our case is,
\begin{eqnarray}
	\label{S00}	\Sigma(V_0,V_0) = \frac{1}{2} \sum_{k=1}^{2} \left[ \left(1+2u^{(0)}_k \right) \ln\left(u^{(0)}_k+\frac{1}{2}\right) + \left(1-2u^{(0)}_k \right) \ln\left(u^{(0)}_k-\frac{1}{2}\right) \right],
\end{eqnarray}
where $u^{(0)}_k,\ k=1,2$ are the symplectic eigenvalues of $V_0$ given by
\begin{eqnarray}
	\label{u01}	u^{(0)}_1 &=& \frac{1}{2} \left(w_{11}^{(0)}+w_{22}^{(0)} +\sqrt{4 w_{12}^{(0)2}+(w_{11}^{(0)}-w_{22}^{(0)})^2}\right)\\
	\label{u02}	u^{(0)}_2 &=& \frac{1}{2} \left(w_{11}^{(0)}+w_{22}^{(0)} -\sqrt{4 w_{12}^{(0)2}+(w_{11}^{(0)}-w_{22}^{(0)})^2}\right).
\end{eqnarray}
The second term in Eq. (\ref{QRE1}), for the case at hand is given by
\begin{eqnarray}
	\label{S01}	\Sigma(V_0,V_1) = \frac{1}{2} \sum_{k=1}^{2} \left[ \left(1+2d_k \right) \ln\left(u_k+\frac{1}{2}\right) + \left(1-2d_k \right) \ln\left(u_k-\frac{1}{2}\right) \right],
\end{eqnarray}
where $u_k$ are the symplectic eigenvalues of $V_1$ and $d_k$ are the diagonal elements (which are doubly degenerate in our case, just as the symplectic spectrum of a CM) of $V_0'=M^T V_0 M$, where $M$ is the symplectic eigenvectors' matrix of $V_1$.
The symplectic eigenvalues of $V_1$ are found to be
\begin{eqnarray}
	\label{u1}	u_1 &=& \frac{1}{2} \left(w_{11}+w_{22} +\sqrt{4 w_{12}^{2}+(w_{11}-w_{22})^2}\right)\\
	\label{u2}	u_2 &=& \frac{1}{2} \left(w_{11}+w_{22} -\sqrt{4 w_{12}^{2}+(w_{11}-w_{22})^2}\right)
\end{eqnarray}
and the symplectic eigenvectors' matrix of $V_1$ is a beam splitter followed by a $2$-mode phase rotation, 
\begin{eqnarray}
	M= R\left(\frac{\theta}{2}+\pi,-\frac{\theta}{2}\right)B(\tau)=
	\begin{pmatrix}
		-\sqrt{\tau}\cos\frac{\theta}{2}&\sqrt{1-\tau}\sin\frac{\theta}{2}&\sqrt{\tau}\sin\frac{\theta}{2}&-\sqrt{1-\tau}\sin\frac{\theta}{2}\\
		\sqrt{1-\tau}\cos\frac{\theta}{2}&\sqrt{\tau}\cos\frac{\theta}{2}&\sqrt{1-\tau}\sin\frac{\theta}{2}&\sqrt{\tau}\sin\frac{\theta}{2}\\
		-\sqrt{\tau}\sin\frac{\theta}{2}&\sqrt{1-\tau}\sin\frac{\theta}{2}&-\sqrt{\tau}\cos\frac{\theta}{2}&\sqrt{1-\tau}\cos\frac{\theta}{2}\\
		-\sqrt{1-\tau}\sin\frac{\theta}{2}&-\sqrt{\tau}\sin\frac{\theta}{2}&\sqrt{1-\tau}\cos\frac{\theta}{2}&\sqrt{\tau}\cos\frac{\theta}{2}
	\end{pmatrix},
\end{eqnarray}
where
\begin{eqnarray}
	\tau = \frac{w_{11}-w_{22}}{2\sqrt{4 w_{12}^2+(w_{11}-w_{22})^2}}+\frac{1}{2}.
\end{eqnarray}
The matrix $M$ can be easily checked to satisfy $MV_1M^T=\textrm{diag}(u_0,u_1,u_0,u_1)$ and $M\Omega M^T=\Omega$, 
where,
\begin{eqnarray}
	\Omega=
	\begin{pmatrix}
		0_{2\times 2} & I_{2\times 2}\\
		-I_{2\times 2} & 0_{2\times 2}
	\end{pmatrix},
\end{eqnarray}
meaning that $M$ is a valid symplectic matrix that diagonalizes (in the symplectic sense) the matrix $V_1$.  
The diagonal elements of $V_0'$ are,
\begin{eqnarray}
	\label{d1}	d_1&=&\frac{w_{11}^{(0)}+w_{22}^{(0)}}{2}+\frac{1}{2\sqrt{4 w_{12}^2+(w_{11}-w_{22})^2}}\left(4w_{12}w_{12}^{(0)}+(w_{11}-w_{22})(w_{11}^{(0)}-w_{22}^{(0)})\right)\\
	\label{d2}	d_2&=&\frac{w_{11}^{(0)}+w_{22}^{(0)}}{2}-\frac{1}{2\sqrt{4 w_{12}^2+(w_{11}-w_{22})^2}}\left(4w_{12}w_{12}^{(0)}+(w_{11}-w_{22})(w_{11}^{(0)}-w_{22}^{(0)})\right).
\end{eqnarray} 
Given the expressions (\ref{u01}), (\ref{u02}), (\ref{u1}), (\ref{u2}), (\ref{d1}), (\ref{d2}), the relative entropy is now known as a function of the $V_0$ and $V_1$ matrix elements through Eqs. (\ref{QRE1}), (\ref{S00}), and (\ref{S01}) (see also Mathematica file QRE.nb).

\subsection{Taylor expansion of the relative entropy}\label{SM:Taylor}
We expand the QRE (\ref{QRE1}) in Taylor series with respect to $\bar{n}_{\textrm{S}}$ at $\bar{n}_{\textrm{S}}=0$
\begin{eqnarray}
	\label{Taylor1}	D(\hat{\rho}_0\|\hat{\rho}_1) \equiv D(\bar{n}_{\textrm{S}}) = D(0) + \frac{\partial D(\bar{n}_{\textrm{S}})}{\partial \bar{n}_{\textrm{S}}}\Bigg|_{\bar{n}_{\textrm{S}}=0} +\frac{1}{2!}\frac{\partial^2 D(\bar{n}_{\textrm{S}})}{\partial \bar{n}_{\textrm{S}}^2}\Bigg|_{\bar{n}_{\textrm{S}}=0} + \frac{1}{3!}\frac{\partial^3 D(\bar{n}_{\textrm{S}})}{\partial \bar{n}_{\textrm{S}}^3}\Bigg|_{\bar{n}_{\textrm{S}}=0}+\ldots
\end{eqnarray}
The $0^{\textrm{th}}$ term in (\ref{Taylor1}) is zero because $D(0)=D(\hat{\rho}_0\|\hat{\rho}_0) = 0$. The $1^{\textrm{st}}$ order term is zero, as the QRE has a minimum at $D(\hat{\rho}_0\|\hat{\rho}_1) = 0$, and the $2^{\textrm{nd}}$ order term is non-negative as the QRE $D(\hat{\rho}_0\|\hat{\rho}_1) \geq 0$, i.e., it has a global minimum at zero. Note that $\Sigma(V_0,V_0)$ of the QRE (\ref{QRE1}) has not a dependence on $\bar{n}_{\textrm{S}}$, therefore one is concerned only with the term $\Sigma(V_0,V_1)$ when taking derivatives on the QRE with respect to $\bar{n}_{\textrm{S}}$.

Having expressed the QRE as a function of the CM elements and consequently as a function of $(\eta_1,\eta_2,\bar{n}_{\textrm{B}_1},\bar{n}_{\textrm{B}_2})$, it is straightforward to derive the $2^{\textrm{nd}}$ and $3^{\textrm{rd}}$ order derivatives. Below, we give a guide on how to derive that the $2^{\textrm{nd}}$ derivative (at $\bar{n}_{\textrm{S}}=0$) is non-negative, that $\mu_c \geq 1$, and that the $3^{\textrm{rd}}$ derivative (at $\bar{n}_{\textrm{S}}=0$) is negative. 
The main tool is to upper and lower bound the $\ln(.)$ \cite{LogBounds} with
\begin{eqnarray}
	\label{logbound} \frac{2x}{x+2} \leq \ln\left(1+x\right) \leq \frac{x}{2}\frac{x+2}{x+1},\ x\geq 0.	
\end{eqnarray}  
The expression of the $2^{\textrm{nd}}$ derivative has the form
\begin{eqnarray}
	c_2=\frac{\partial^2 D(\bar{n}_{\textrm{S}})}{\partial \bar{n}_{\textrm{S}}^2}\Bigg|_{\bar{n}_{\textrm{S}}=0}= T_1+T_2 \ln\left(1+x\right),
\end{eqnarray} 
where $T_1$, $T_2>0$, and $x>0$ are functions of $(\eta_1,\eta_2,\bar{n}_{\textrm{B}_1},\bar{n}_{\textrm{B}_2})$. Since $T_2>0$, we use the lower bound of (\ref{logbound}), $c_2 \geq T_1+T_2\frac{2x}{x+2}$, from which expression it can shown that $c_2\geq 0$.

To prove that $\mu_c=c_{\textrm{ASE}}/c_{\textrm{coh}} \geq 1 \Rightarrow c_{\textrm{ASE}}-c_{\textrm{coh}} \geq 0 $, where $c_{\textrm{ASE}} = [1+2\bar{n}_{\textrm{B}_\textrm{eff}}(1-\eta_\textrm{eff})]\sqrt{c_2}/(16\eta_{\textrm{eff}})$, we use a similar technique. We eliminate the $\ln(.)$ in $c_2$ using again the lower bound of (\ref{logbound}, a technique which gives a sufficient, but not necessary, condition. However, we get indeed $\mu_c \geq 1$.

The $3^{\textrm{rd}}$ derivative of the QRE has the form,
\begin{eqnarray}
	c_3=\frac{\partial^3 D(\bar{n}_{\textrm{S}})}{\partial \bar{n}_{\textrm{S}}^3}\Bigg|_{\bar{n}_{\textrm{S}}=0}= P_1+P_2 +P_3\ln\left(1+x\right),
\end{eqnarray} 
where $P_1$, $P_2$, $P_3$, and $x>0$ are functions of $(\eta_1,\eta_2,\bar{n}_{\textrm{B}_1},\bar{n}_{\textrm{B}_2})$. In this case, $P_3$ can be positive or negative. Therefore, one has to take cases when eliminating the $\ln(.)$, which again gives a sufficient, but not necessary, condition. When $P_3>0$, we use the upper bound of (\ref{logbound}), $c_3 \leq P_1+P_2+P_3\frac{x}{2}\frac{x+2}{x+1}$, while when $P_3<0$, we use the lower bound of (\ref{logbound}), $c_3 \leq P_1+P_2+P_3\frac{2x}{x+2}$. For both cases it can shown that $c_3\leq 0$.

The case where $\bar{n}_{\textrm{B}_1}=\bar{n}_{\textrm{B}_2}=\bar{n}_{\textrm{B}}$ yields simple results,
\begin{eqnarray}
	D(\hat{\rho}_0\|\hat{\rho}_1)_{(\bar{n}_{\textrm{B}})} &=&\eta_{\textrm{eff}} \bar{n}_{\textrm{B}}\ln\left[\frac{\eta_{\textrm{eff}} \bar{n}_\textrm{B}}{1+\eta_{\textrm{eff}} \bar{n}_\textrm{B}}\frac{1+(1-\eta_{\textrm{eff}})\bar{n}_{\textrm{S}}+\eta_{\textrm{eff}}\bar{n}_{\textrm{B}}}{(1-\eta_{\textrm{eff}})\bar{n}_{\textrm{S}}+\eta_{\textrm{eff}}\bar{n}_{\textrm{B}}}\right]+\ln\left[\frac{1+(1-\eta_{\textrm{eff}})\bar{n}_{\textrm{S}}+\eta_{\textrm{eff}}\bar{n}_{\textrm{B}}}{(1-\eta_{\textrm{eff}})\bar{n}_{\textrm{S}}+\eta_{\textrm{eff}}\bar{n}_{\textrm{B}}}\right],\\
	c_{2\ (\bar{n}_{\textrm{B}})} &=& \frac{(1-\eta_{\textrm{eff}})^2}{\eta_{\textrm{eff}}\bar{n}_{\textrm{B}}(1+\eta_{\textrm{eff}}\bar{n}_{\textrm{B}})}>0,\\
	\lambda_c &=&1,\\
	c_{3\ (\bar{n}_{\textrm{B}})}  &=& -\frac{2(1-\eta_{\textrm{eff}})^3(1+2\eta_{\textrm{eff}}\bar{n}_{\textrm{B}})}{\eta_{\textrm{eff}}^2\bar{n}_{\textrm{B}}^2(1+\eta_{\textrm{eff}}\bar{n}_{\textrm{B}})^2}<0. 	
\end{eqnarray}  
Details can be found in Mathematica file QRE.nb.

\subsection{Effective channel}\label{SM:Effective}
A covariance matrix $V_1$ under a thermal channel $C(\bar{n}_{\textrm{B}},\eta)$ \cite{Holevo2008} is transformed as
\begin{eqnarray}
	V_2 = X V_1 X^T +Y,
\end{eqnarray}   
where,
\begin{eqnarray}
	\label{eq:X}	X&=&\sqrt{\eta} I\\
	\label{eq:Y}	Y&=&(1-\eta)\left(\bar{n}_{\textrm{B}}-\frac{1}{2}\right) I.
\end{eqnarray}
Alice's sensing arm goes through a thermal channel $C_1(\bar{n}_{\textrm{B}_1},\eta_1)$, a phase shift $R(\theta)$ which is a symplectic orthogonal matrix, and another thermal channel $C_2(\bar{n}_{\textrm{B}_2},\eta_2)$. Therefore, the sensing arm will see the transformation,
\begin{eqnarray}
	\label{eq:XA}	\tilde{X} &=& X_2 R(\theta) X_1\\
	\label{eq:YA}	\tilde{Y} &=& X_2 R(\theta) Y_1 R^T(\theta) X_2^T + Y_2,
\end{eqnarray}
Where $X_1, Y_1$ and $X_2, Y_2$ correspond to $C_1$ and $C_2$ respectively. Using Eqs. (\ref{eq:X}), (\ref{eq:Y}), and since the matrices $X_1, X_2, Y_1, Y_2$ are proportional to the identity matrix $I$ they commute with $R(\theta)$, Eqs. (\ref{eq:XA}) and (\ref{eq:YA}) give,
\begin{eqnarray}
	\label{eq:Xeff}	\tilde{X} &=& \sqrt{\eta_{\textrm{eff}}} I\\
	\label{eq:Yeff}	\tilde{Y} &=& (1-\eta_{\textrm{eff}}) \left(\bar{n}_{\textrm{B}_{\textrm{eff}}}+\frac{1}{2}\right)I,
\end{eqnarray}
where,
\begin{eqnarray}
	\label{eq:etaeff}	\eta_{\textrm{eff}} &=& \eta_1\eta_2\\
	\label{eq:nbeff}	 \bar{n}_{\textrm{B}_{\textrm{eff}}} &=& \frac{(1-\eta_1)\eta_2 \bar{n}_{\textrm{B}_1}+(1-\eta_2)\bar{n}_{\textrm{B}_2} }{1-\eta_1\eta_2}.
\end{eqnarray}
Therefore, Alice's sensing arm sees only one effective thermal channel with transmittance $\eta_{\textrm{eff}}$, mean thermal photon number $\bar{n}_{\textrm{B}_{\textrm{eff}}}$, and a phase shift $R(\theta)$. The phase shift $R(\theta)$ can be put before or after the thermal channel as it commutes with $\tilde{X}$ and $\tilde{Y}$ and $R(\theta)$ is orthogonal. The matrix elements $a_{11}$ and $a_{12}$ in Alice's CM can now be written,
\begin{eqnarray}
	\label{eq:a11eff} a_{11} &=& (1-\eta_{\textrm{eff}}) \bar{n}_{\textrm{B}_{\textrm{eff}}} + \eta_{\textrm{eff}} \bar{n}_{\textrm{S}} +\frac{1}{2},\\
	\label{eq:a12eff} a_{12} &=& -\sqrt{\eta_{\textrm{eff}} \bar{n}_{\textrm{S}} \bar{n}_{\textrm{LO}}} \cos\theta.
\end{eqnarray}

\subsection{Heterodyne detection}\label{SM:Het}
\begin{figure}
	\centering
	\includegraphics[width=1\linewidth]{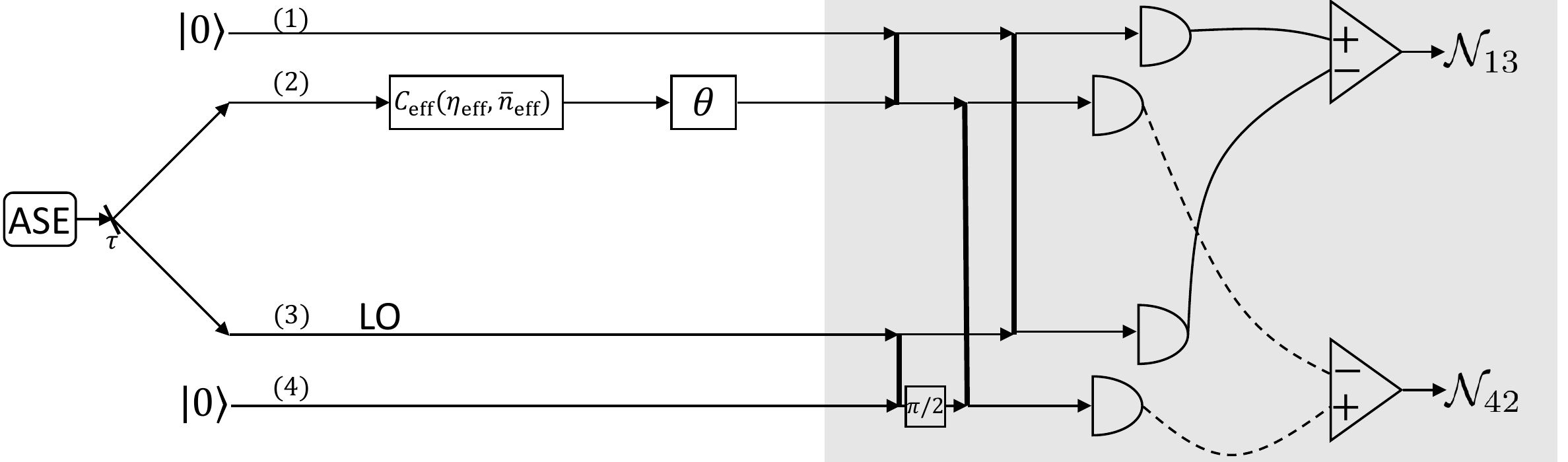}
	\caption{The heterodyne detection takes place in the gray area. Bold vertical lines represent balanced beam splitters. Alice splits her local oscillator and the sensing arm into two equal beams each. On the lower local oscillator a phase $\pi/2$ is applied. Then, modes $(1)$ and $(2)$ of the sensing arm are mixed with the modes $(3)$ and $(4)$ respectively. Finally, a homodyne is performed on the couples of modes $(1),(3)$ and $(2),(4)$ using photon number resolution detectors.}
	\label{fig:heterodyne}
\end{figure}

We will consider dual homodyne detection on Alice's system as in Fig.~\ref{fig:heterodyne}, which is an implementation of heterodyne detection. We want to calculate the mean values $\mu_{13}$, $\mu_{42}$ and the variances $\sigma^2_{13}$, $\sigma^2_{42}$ of the Gaussian distributions $\mathcal{N}(\mu_{13},\sigma^2_{13}),\ \mathcal{N}(\mu_{42},\sigma^2_{42})$ respectively. The aforementioned Gaussian distributions correspond to the observables $\hat{n}_1-\hat{n}_3$ and $\hat{n}_4-\hat{n}_2$ in the following manner, 
\begin{eqnarray}
	\label{mu13}	\mu_{13} &=&\langle \hat{n}_1-\hat{n}_3 \rangle\\
	\label{mu42}	\mu_{42} &=& \langle \hat{n}_4-\hat{n}_2 \rangle\\
	\label{s13}	\sigma^2_{13} &=&\langle (\hat{n}_1-\hat{n}_3)^2 \rangle = \langle \hat{n}_1^2 \rangle - \langle \hat{n}_1 \rangle^2 + \langle \hat{n}_3^2 \rangle - \langle \hat{n}_3 \rangle^2 - 2 \left(\langle\hat{n}_1\hat{n}_3\rangle-\langle\hat{n}_1\rangle\langle\hat{n}_3\rangle\right) \\
	\label{s42}	\sigma^2_{42} &=&\langle (\hat{n}_4-\hat{n}_4)^2 \rangle = \langle \hat{n}_4^2 \rangle - \langle \hat{n}_4 \rangle^2 + \langle \hat{n}_2^2 \rangle - \langle \hat{n}_2 \rangle^2 - 2 \left(\langle\hat{n}_4\hat{n}_2\rangle-\langle\hat{n}_4\rangle\langle\hat{n}_2\rangle\right).
\end{eqnarray}
Therefore, we need to calculate the expectation values in Eqs.~(\ref{mu13}), (\ref{mu42}), (\ref{s13}), and (\ref{s42}). To this end, we calculate the evolution of the CM for the setup of Fig.~\ref{fig:heterodyne}. The CM $V_{\textrm{in}}$ just before the heterodyne detection is given by the Eq. (\ref{VA}) and by taking into account that the CM for vacuum is $I/2$,
\begin{eqnarray}
	\label{Vin}	V_{\textrm{in}}=
	\begin{pmatrix}
		\frac{1}{2} & 0 & 0 & 0 & 0 & 0 & 0 & 0\\
		0 & a_{11} & -a_{12} \cos\theta & 0 & 0 & 0 & a_{12}\sin\theta & 0\\
		0 & -a_{12} \cos\theta & a_{22} & 0 & 0 & -a_{12} \sin\theta & 0 & 0\\
		0 & 0 & 0 & \frac{1}{2} & 0 & 0 & 0 & 0\\
		0 & 0 & 0 & 0 & \frac{1}{2} & 0 & 0 & 0\\
		0 & 0 & -a_{12} \sin\theta & 0 & 0 & a_{11} & -a_{12} \cos\theta & 0\\
		0 & a_{12} \sin\theta & 0 & 0 & 0 & -a_{12} \cos\theta & a_{22} & 0\\
		0 & 0 & 0 & 0 & 0 & 0 & 0 & \frac{1}{2}
	\end{pmatrix}.
\end{eqnarray}
After the application of all symplectic matrices as described in Fig.~\ref{fig:heterodyne}, the CM just before the photon number resolution detectors is,
\begin{eqnarray}
	\label{Vhet} V_{\textrm{het}}&=&
	\begin{pmatrix}
		A & B \\
		B^T & A\\
	\end{pmatrix},\\
	A&=&
	\begin{pmatrix}
		A_{11} & A_{12} & A_{13} & A_{14}\\
		A_{12} & A_{22} & A_{23} & A_{24}\\
		A_{13} & A_{23} & A_{33} & A_{34}\\
		A_{14} & A_{24} & A_{34} & A_{44}
	\end{pmatrix},
\end{eqnarray}
\begin{eqnarray}
	B&=&
	\begin{pmatrix}
		0 & B_{12} & B_{13} & B_{14}\\
		-B_{12} & 0 & B_{23} & B_{24}\\
		-B_{13} & -B_{23} & 0 & B_{34}\\
		-B_{14} & -B_{24} & -B_{34} & 0
	\end{pmatrix},
\end{eqnarray}
\begin{eqnarray}
	A_{11}&=& \frac{1}{4} (1 + a_{11} + a_{22} - 2 a_{12} \cos\theta)\\
	A_{12}&=& \frac{1}{8} [1 - 2 a_{11} + 2 a_{12}(\cos\theta - \sin\theta)]\\
	A_{13}&=&  \frac{a_{11}-a_{22}}{4}\\
	A_{14}&=& \frac{1}{8} [1 - 2 a_{11} + 2 a_{12}(\cos\theta + \sin\theta)]\\
	A_{22}&=& \frac{1}{4} (1 + a_{11} + a_{22} + 2 a_{12} \sin\theta)\\
	A_{23}&=& \frac{1}{8} [1 - 2 a_{11} - 2 a_{12} (\cos\theta + \sin\theta)]\\
	A_{24}&=& \frac{a_{11}-a_{22}}{4}\\
	A_{33}&=&  \frac{1}{4} (1 + a_{11} + a_{22} + 2 a_{12} \cos\theta)\\
	A_{34}&=& \frac{1}{8} [1 - 2 a_{11} - 2 a_{12}( \cos\theta - \sin\theta)] \\
	A_{44}&=& \frac{1}{4} (1 + a_{11} + a_{22} - 2 a_{12} \sin\theta),
\end{eqnarray}
\begin{eqnarray}
	B_{12}&=& -\frac{1}{8} [1 - 2 a_{22} + 2 a_{12} (\cos\theta - \sin\theta)]\\
	B_{13}&=& -\frac{1}{2} a_{12} \sin\theta\\
	B_{14}&=& \frac{1}{8} [1 - 2 a_{22} + 2 a_{12} (\cos\theta + \sin\theta)]\\
	B_{23}&=& -\frac{1}{8} [1 - 2 a_{22} - 2 a_{12} (\cos\theta + \sin\theta)]\\
	B_{24}&=& -\frac{1}{2} a_{12} \cos\theta\\
	B_{34}&=& -\frac{1}{8} [1 - 2 a_{22} - 2 a_{12} (\cos\theta - \sin\theta)].
\end{eqnarray}
Using the the CM of Eq. (\ref{Vhet}), we can construct the Wigner function,
\begin{eqnarray}
	\label{Wigner} W(\bm{r}) = \frac{1}{(2\pi)^4\sqrt{\det V_{\textrm{het}}}} \int d^2\bm{r} \exp\left(-\frac{1}{2} \bm{r}^T V_{\textrm{het}}^{-1}\bm{r}\right)
\end{eqnarray}
where $\bm{r}=(q_1,\ldots,q_4,p_1,\ldots,p_4)$ is the phase space vector ($q_i,p_i$ refer to position and momentum respectively of the $i$-th mode, also note that we work with $\hbar=1$), and $d^2\bm{r} = dq_1\ldots dq_4 dp_1\ldots dp_4$. Utilizing the Wigner function we can calculate mean values in the symmetric ordering,
\begin{eqnarray}
	\langle : \hat{n}_i \hat{n}_j : \rangle = \int d^2\bm{r} W(\bm{r}) \frac{(q_i^2+p_i^2)(q_j^2+p_j^2)}{4}.
\end{eqnarray}
Utilizing Wick-Isserlis theorem and the commutation relation $[\hat{a}_i,\hat{a}_j^\dagger]=\delta_{ij}$, for example, 
\begin{eqnarray}
	\langle : \hat{n}_i \hat{n}_j : \rangle = \langle  \hat{n}_i \hat{n}_j  \rangle+\frac{1}{2}\langle\hat{n}_i\rangle+\frac{1}{2}\langle\hat{n}_j\rangle-\frac{1}{4},
\end{eqnarray}
we find
\begin{eqnarray}
	\label{n1}	\langle\hat{n}_1\rangle &=& \frac{1}{4} (a_{11} + a_{22} - 2 a_{12} \cos\theta-1)\\
	\label{n2}	\langle\hat{n}_2\rangle &=&\frac{1}{4} (a_{11} + a_{22} + 2 a_{12} \sin\theta-1)\\
	\label{n3}	\langle\hat{n}_3\rangle &=&\frac{1}{4} (a_{11} + a_{22} + 2 a_{12} \cos\theta-1)\\
	\label{n4}	\langle\hat{n}_4\rangle &=&\frac{1}{4} (a_{11} + a_{22} - 2 a_{12} \sin\theta-1)\\
	\label{n12}	\langle\hat{n}_1^2 \rangle &=& \frac{1}{8} (a_{11} + a_{22} - 2 a_{12} \cos\theta-1) ( a_{11} + a_{22} - 2 a_{12} \cos\theta+1)\\
	\label{n22}	\langle\hat{n}_2^2 \rangle &=& \frac{1}{8} (a_{11} + a_{22} + 2 a_{12} \sin\theta-1) ( a_{11} + a_{22} + 2 a_{12} \sin\theta+1)\\
	\label{n32}	\langle\hat{n}_3^2 \rangle &=& \frac{1}{8} (a_{11} + a_{22} + 2 a_{12} \cos\theta-1) ( a_{11} + a_{22} + 2 a_{12} \cos\theta+1) \\
	\label{n42}	\langle\hat{n}_4^2 \rangle &=& \frac{1}{8} (a_{11} + a_{22} - 2 a_{12} \sin\theta-1) ( a_{11} + a_{22} - 2 a_{12} \sin\theta+1)\\
	\label{n1n3}	\langle\hat{n}_1 \hat{n}_3 \rangle &=& \frac{1}{16} [1 + 2 (a_{11}-1) a_{11} + 2 (a_{22}-1) a_{22} - 4 a_{12}^2 \cos 2\theta] \\
	\label{n4n2}	\langle\hat{n}_4 \hat{n}_2 \rangle &=& \frac{1}{16} [1 + 2 (a_{11}-1) a_{11} + 2 (a_{22}-1) a_{22} + 4 a_{12}^2 \cos 2\theta].
\end{eqnarray}
From Eqs. (\ref{eq:a22}), (\ref{eq:a11eff}), (\ref{eq:a12eff}), (\ref{mu13}), (\ref{mu42}), (\ref{s13}), (\ref{s42}), and (\ref{n1}) to (\ref{n4n2}), we find
\begin{eqnarray}
	\mu_{13} &=& \sqrt{\eta_{\textrm{eff}} \bar{n}_{\textrm{S}} \bar{n}_{\textrm{LO}} } \cos\theta\\
	\mu_{42} &=& \sqrt{\eta_{\textrm{eff}} \bar{n}_{\textrm{S}} \bar{n}_{\textrm{LO}} } \sin\theta\\
	\sigma_{13}^2 &=&\frac{1}{2} [\bar{n}_{\textrm{LO}} + \bar{n}_{\textrm{eff}} (1 - \eta_{\textrm{eff}}) + \bar{n}_{\textrm{eff}} (1 - \eta_{\textrm{eff}}) \bar{n}_{\textrm{LO}} + \bar{n}_{\textrm{S}} \eta_{\textrm{eff}}] + \bar{n}_{\textrm{LO}} \bar{n}_{\textrm{S}} \eta_{\textrm{eff}} \cos^2\theta\\
	\sigma_{42}^2 &=& \frac{1}{2} [\bar{n}_{\textrm{LO}} + \bar{n}_{\textrm{eff}} (1 - \eta_{\textrm{eff}}) + \bar{n}_{\textrm{eff}} (1 - \eta_{\textrm{eff}}) \bar{n}_{\textrm{LO}} + \bar{n}_{\textrm{S}} \eta_{\textrm{eff}}] + \bar{n}_{\textrm{LO}} \bar{n}_{\textrm{S}} \eta_{\textrm{eff}} \sin^2\theta.
\end{eqnarray}
We normalize the mean values with $k=\sqrt{\eta_{\textrm{eff}} \bar{n}_{\textrm{S}} \bar{n}_{\textrm{LO}} }$ and consequently the variances with $k^2$, and by taking the limit of a very strong local oscillator we get,
\begin{eqnarray}
	\mu_{1} &=& \frac{\mu_{13}}{k} = \cos\theta\\
	\mu_{2} &=& \frac{\mu_{42}}{k} = \sin\theta\\
	\sigma_{1}^2 &=&\lim\limits_{\bar{n}_{\textrm{LO}}\rightarrow \infty} \frac{\sigma_{13}}{k^2} =\sigma^2+\cos^2\theta\\
	\sigma_{2}^2 &=& \lim\limits_{\bar{n}_{\textrm{LO}}\rightarrow \infty} \frac{\sigma_{42}}{k^2} =\sigma^2+\sin^2\theta,
\end{eqnarray}
where
\begin{eqnarray}
	\sigma^2 = \frac{1+(1-\eta_{\textrm{eff}})\bar{n}_{\textrm{B}_{\textrm{eff}}}}{2 \eta_{\textrm{eff}} \bar{n}_{\textrm{S}}}.
\end{eqnarray}
Alice is trying to determine the mean values of the distributions $\mathcal{N}(\cos\theta, \sigma^2+\cos^2\theta)$ and $\mathcal{N}(\sin\theta, \sigma^2+\sin^2\theta)$ by using the channel $n$ times, i.e, by performing $n$ measurements. Then the distributions of the mean values will be  $\mathcal{N}(\cos\theta, \sigma^2/n+\cos^2\theta/n)$ and $\mathcal{N}(\sin\theta, \sigma^2/n+\sin^2\theta/n)$, were we have used the fact that independent Gaussian variables are additive. We set,
\begin{eqnarray}
	\sigma^2_{\textrm{het}}=\frac{\sigma^2}{n}=\frac{1+(1-\eta_{\textrm{eff}})\bar{n}_{\textrm{B}_{\textrm{eff}}}}{2 n \eta_{\textrm{eff}} \bar{n}_{\textrm{S}}}
\end{eqnarray}
and by imposing the covertness condition $\bar{n}_\textrm{S} =(4 \epsilon)/[\sqrt{c_2}\sqrt{n}]$ we get,
\begin{eqnarray}
	\label{sigmaHet}\sigma^2_{\textrm{het}} = \frac{\tilde{c}_{\textrm{het}}}{\epsilon\sqrt{n}},
\end{eqnarray}
where $\tilde{c}_{\textrm{het}}=[1+\bar{n}_{\textrm{B}_\textrm{eff}}(1-\eta_\textrm{eff})]\sqrt{c_2}/(8\eta_{\textrm{eff}})$.
Therefore the two distributions now are  $\mathcal{N}(\cos\theta,  \tilde{c}_{\textrm{het}}/(\epsilon\sqrt{n})+\cos^2\theta/n)$ and $\mathcal{N}(\sin\theta,  \tilde{c}_{\textrm{het}}/(\epsilon\sqrt{n})+\sin^2\theta/n)$.
In the limit of $n \gg 1$ (a limit which is justified by the large bandwidth of the ASE source), since $\sin^2\theta \leq 1$ and $\cos^2\theta \leq 1$ the term that scales as $1/n$ will be insignificant, therefore the approach in Sec.\ref{Sec:Estimator} and in \cite{bash2017isit} (Appendix C therein) applies here as well and we conclude that the mean square error of the estimator $\hat{\theta}_{\textrm{het},n}=\arctan\left((\sin\theta+\sigma^2_\textrm{het})/(\cos\theta+\sigma^2_\textrm{het})\right)$ scales as $\langle (\theta-\hat{\theta}_{\textrm{het},n})^2\rangle =\mathcal{O}(1/\sqrt{n})$.

\subsection{Proof that $\pmb{\left\langle (\theta - {\hat \theta}_{\rm het})^2 \right\rangle \leq \sigma^2_{\rm het}+\mathcal{O}\left(\frac{1}{n}\right)}$ }\label{Sec:Estimator}
Consider the following estimator for $\theta$,
\begin{eqnarray}
	\label{eq:estimator_het}\hat{\theta}_{\rm het}&=&\tan^{-1}\left(\frac{\sin(\theta)+\frac{1}{n}\sum_{i=1}^n Z^{\rm (Q,N)}_i}{\cos(\theta)+\frac{1}{n}\sum_{i=1}^n Z^{\rm (I,N)}_i}\right)\\
	&=&\tan^{-1}\left(\frac{\sin(\theta)+Z^{\rm (Q)}}{\cos(\theta)+Z^{\rm (I)}}\right),
\end{eqnarray}
where the random variables representing in-phase (I) and quadrature (Q) noise $Z^{\rm (I)},\ Z^{\rm (Q)}$ respectively, obey the Gaussian distributions $Z^{\rm (I)}\sim\mathcal{N}(0,\sigma^2_{\rm het})$ and $Z^{\rm (Q)}\sim\mathcal{N}(0,\sigma^2_{\rm het})$, $Z^{\rm (I,N)}\sim\mathcal{N}(0,\sigma^2+\cos^2\theta)$ and $Z^{\rm (Q,N)}\sim\mathcal{N}(0,\sigma^2+\sin^2\theta)$. The distributions are indeed Gaussian for $n\gg 1$ and by applying the central limit theorem. The variance $\sigma^2_{\rm het}$ is given in Eq. (\ref{sigmaHet}). The MSE is:
\begin{align}
	\left\langle (\theta - {\hat \theta}_{\rm het})^2 \right\rangle &=\left\langle\left(\theta-\tan^{-1}\left(\frac{\sin(\theta)+Z^{\rm (Q)}}{\cos(\theta)+Z^{\rm (I)}}\right)\right)^2 \right\rangle\\
	\label{eq:sub_polar}&=\left\langle\left(\theta-\tan^{-1}\left(\frac{\sin(\theta)+R\cos(\varphi)}{\cos(\theta)+R\sin(\varphi)}\right)\right)^2 \right\rangle
\end{align}
where in \eqref{eq:sub_polar} we use circular symmetry of the two-dimensional added white Gaussian noise (AWGN) to change from the rectangular to polar coordinate system.
Thus, the radius is distributed as a Rayleigh random variable
$R\sim\text{Rayleigh}(\sigma^2_{\rm het})$
while the angle is distributed uniformly $\varphi\sim\mathcal{U}([0,2\pi])$.
Now, the Taylor series expansion of
$\tan^{-1}\left(\frac{\sin(\theta)+r\cos(\varphi)}{\cos(\theta)+r\sin(\varphi)}\right)$ around $r=0$ is:
\begin{align}
	\tan^{-1}\left(\frac{\sin(\theta)+r\cos(\varphi)}{\cos(\theta)+r\sin(\varphi)}\right)&=\theta+r\cos(\theta+\varphi)-\frac{r^2}{2}\sin(2(\theta+\varphi))-\frac{r^3}{3}\cos(3(\theta+\varphi))+\frac{r^4}{4}\sin(4(\theta+\varphi))\nonumber\\
	&\phantom{=}+\frac{r^5}{5}\cos(5(\theta+\varphi))-\frac{r^6}{6}\sin(6(\theta+\varphi))-\frac{r^7}{7}\cos(7(\theta+\varphi))+\frac{r^8}{8}\sin(8(\theta+\varphi))\nonumber\\
	\label{eq:taylor}&\phantom{=}+\ldots\\
	\label{eq:taylor_ub}&\leq\theta+r\cos(\theta+\varphi)+\sum_{i=2}^\infty \frac{r^i}{i}\\
	\label{eq:taylor_log}&=\theta+r\cos(\theta+\varphi)-(\log(1-r)+r)~\text{provided}~0\leq r <1,
\end{align}
where the upper bound in \eqref{eq:taylor_ub} is because 
$\sin(x),\cos(x)\in[-1,1]$.
While this demonstrates the convergence of the Taylor series for 
$r<1$, the $n^{\mathrm{th}}$ root test shows that the Taylor series in 
\eqref{eq:taylor} does 
not converge for $r>1$ (the series converges for $r=1$ by the alternating 
series test, however, this is a zero-probability event).
However, since $\tan^{-1}(x)\in \left[-\frac{\pi}{2},\frac{\pi}{2}\right]$ and 
$\theta\in \left(-\frac{\pi}{2},\frac{\pi}{2}\right)$, for any $r$ and 
$\varphi$, 
\begin{align}
	\label{eq:arctan_trivial_bound}\left|\tan^{-1}\left(\frac{\sin(\theta)+r\cos(\varphi)}{\cos(\theta)+r\sin(\varphi)}\right)-\theta\right|\leq \pi.
\end{align}
Therefore, using the Taylor series expansion of $\log(1-x)$ around $x=0$ in
\eqref{eq:taylor_log}, and \eqref{eq:arctan_trivial_bound}, it is 
straightforward to show there exist constants $a\in(0,1)$ and $b>0$ such that:
\begin{align}
	\label{eq:arctan_ub}\left|\tan^{-1}\left(\frac{\sin(\theta)+r\cos(\varphi)}{\cos(\theta)+r\sin(\varphi)}\right)-\theta\right|\leq \left\{\begin{array}{ll}r\cos(\theta+\varphi)+br^2&\text{if}~r\leq a\\\pi&\text{otherwise}\end{array}\right.
\end{align}
We can use \eqref{eq:arctan_ub} to upper bound the MSE,
\begin{align}
	\left\langle (\theta - {\hat \theta}_{\rm het})^2 \right\rangle &\leq\frac{1}{2\pi}\int_0^{2\pi}\left(\int_0^a(r\cos(\theta+\varphi)+br^2)^2\frac{re^{-x^2/2\sigma^2_{\rm het}}}{\sigma^2_{\rm het}}\mathrm{d}r+\int_a^\infty\pi^2\frac{re^{-x^2/2\sigma^2_{\rm het}}}{\sigma^2_{\rm het}}\mathrm{d}r\right)\mathrm{d}\varphi\\
	&=\sigma^2_{\rm het}+8b^2\sigma^4_{\rm het}-\frac{1}{2}e^{-\frac{a^2}{2\sigma^2_{\rm het}}}\left(2a^4b^2+a^2(1+8c^2\sigma^2_{\rm het})+2\sigma^2_{\rm het}(1+8c^2\sigma^2_{\rm het})-2\pi^2\right)\\
	\label{eq:het_mse_approx}&=\sigma^2_{\rm het}+\mathcal{O}(\sigma^4_{\rm het})\\
	&=\sigma^2_{\rm het}+\mathcal{O}\left(\frac{1}{n}\right).
\end{align}

\subsection{Quantum Fisher information}\label{SM:QFI}
The final CM describing Alice's state is
\begin{eqnarray}
	V_A =
	\begin{pmatrix}
		a_{11} & -a_{12}\cos\theta & 0 & a_{12} \sin\theta \\
		-a_{12}\cos\theta & a_{22} & -a_{12} \sin\theta & 0 \\
		0 & -a_{12} \sin\theta & a_{11} & -a_{12}\cos\theta \\
		a_{12} \sin\theta & 0 & -a_{12} \cos\theta & a_{22}
	\end{pmatrix}
\end{eqnarray}
The QFI for the parameter $\theta$ can be calculated using the formalism developed in \cite{Marian2012,Banchi2015}. One first must derive the quantum fidelity
\begin{eqnarray}
	\label{Fidelity}\mathcal{F}(\omega)=\frac{1}{\sqrt{\sqrt{\Gamma}+\sqrt{\Lambda}-\sqrt{(\sqrt{\Gamma}+\sqrt{\Lambda})^2-\Delta}}}
\end{eqnarray}
between the two-mode states with CM $V_A \equiv V_A(\theta)$ and $V_A(\theta+\omega)$ and zero first moments. Equation (\ref{Fidelity}) is based on the symplectic invariants,
\begin{eqnarray}
	\label{Delta}\Delta&=&\det\left(V_A(\theta)+V_A(\theta+\omega)\right) \geq 1\\
	\label{Gamma}\Gamma&=&16 \det\left(\Omega V_A(\theta) \Omega V_A(\theta+\omega)-\frac{I_{4\times 4}}{4} \right) \geq \Delta\\
	\label{Lambda}\Lambda&=&16 \det\left(V_A(\theta)+\frac{i}{2}\Omega\right) \det\left(V_A(\theta+\omega)+\frac{i}{2}\Omega\right) \geq 0
\end{eqnarray}
where,
\begin{eqnarray}
	\Omega=
	\begin{pmatrix}
		0_{2\times 2} & I_{2\times 2}\\
		-I_{2\times 2} & 0_{2\times 2}
	\end{pmatrix}.
\end{eqnarray}
The QFI for the parameter $\theta$ $F_A'$ is given by the second derivative,
\begin{eqnarray}
	\label{QFIA} F_A'=-4 \frac{d^2 \mathcal{F}(\omega)}{d\omega^2}\Bigg|_{\omega=0}= \frac{8 a_{12}^2}{4(a_{11}a_{22}-a_{12}^2)-1}=\frac{8 a_{12}^2}{4\sqrt{\det V_A}-1}.
\end{eqnarray}
Substituting Eqs. (\ref{eq:a11}), (\ref{eq:a12}), (\ref{eq:a22}), (\ref{eq:etaeff}), and (\ref{eq:nbeff}) into Eq. (\ref{QFIA}) we get
\begin{eqnarray}
	\label{QFIA2} F_A' = \frac{4 \bar{n}_{\textrm{LO}} \bar{n}_\textrm{S} \eta_\textrm{eff}}{\bar{n}_{\textrm{LO}}+(1-\eta_\textrm{eff})\bar{n}_{\textrm{B}_\textrm{eff}}(1+2\bar{n}_{\textrm{LO}})+\eta_\textrm{eff} \bar{n}_\textrm{S}}
\end{eqnarray}
and by taking the limit $\bar{n}_{\textrm{LO}}\rightarrow \infty$ we get,
\begin{eqnarray}
	F_A = \lim\limits_{\bar{n}_{\textrm{LO}}\rightarrow \infty} F_A' = \frac{4 \bar{n}_\textrm{S} \eta_{\textrm{eff}} }{1+2\bar{n}_{\textrm{B}_\textrm{eff}}(1-\eta_\textrm{eff})}.
\end{eqnarray}

\end{document}